\documentclass[pra,twocolumn]{revtex4-1}
\usepackage{amsmath,amssymb,mathrsfs}

\usepackage{graphicx,import}
\usepackage[usenames,dvipsnames]{color}
\usepackage[colorlinks=true,linkcolor=blue,citecolor=BrickRed,urlcolor=blue]{hyperref}

\usepackage{circuitikz}

\newcommand{\La}{\mathscr{L}}

\def\be{\begin{equation}}
\def\ee{\end{equation}}
\def\bea{\begin{eqnarray}}
\def\eea{\end{eqnarray}}

\def\v#1{\mathbf{#1}}

\begin{document}
%\tableofcontents

%\begin{frontmatter}
\title{Topological circuits of inductors and capacitors}
\author{Erhai Zhao}
\address{Department of Physics and Astronomy,
  George Mason University, Fairfax, Virginia}

\begin{abstract}
Alternating current (ac) circuits can have electromagnetic edge modes protected by symmetries, analogous to
topological band insulators or semimetals. How to make such a topological circuit?
This paper illustrates a particular design idea by analyzing a series
of topological circuits consisting purely of inductors (L) and capacitors (C) connected to each other by wires
to form periodic lattices. All the examples are treated using a unifying approach based on Lagrangians
and the dynamical $H$-matrix.
First, the building blocks and permutation wiring
are introduced using simple circuits in one dimension, the SSH transmission line and a braided ladder analogous to the ice-tray model also known as the $\pi$-flux ladder.
Then, more general building blocks (loops and stars) and wiring schemes ($m$-shifts) are introduced. The key concepts of emergent pseudo-spin
degrees of freedom and synthetic gauge fields are discussed, and the connection to quantum lattice Hamiltonians is clarified.
A diagrammatic notation is introduced to simplify the design and presentation of more complicated circuits. These building blocks
are then used to construct topological circuits in higher dimensions. The examples include
the circuit analog of Haldane's Chern insulator in two dimensions and
quantum Hall insulator in four dimensions featuring finite second Chern numbers.
The topological invariants and symmetry protection of the edge modes are discussed based on the $H$-matrix.
 \end{abstract}
 
% \begin{keyword}
%Linear circuits \sep topological invariants \sep edge states  
 %\end{keyword}
%\pacs{}
%\end{frontmatter}

\maketitle

\section{Introduction}
Classical many-body systems including mechanical \cite{kane2014topological,stenull2016topological,nash2015topological,susstrunk2015observation,huber2016topological,chen2016topological,susstrunk2016classification}, acoustic \cite{prodan2009topological,peano2015topological,xiao2015geometric,wang2015topological,PhysRevLett.114.114301}, and photonic \cite{lu2015experimental,lu2014topological} metamaterials are receiving a renewed interest regarding their topological properties. In light of the active experimental and theoretical developments in these areas, it is desirable to have simple toy models to examine their nontrivial band topologies and the corresponding edge modes. Hopefully, the toy model can be built and tinkered with by non-experts using cheap, easily available parts. And ideally, the model is also  intuitive, i.e., based on familiar concepts such as coupled harmonic oscillators. The goal of this paper is to present and analyze a few toy models that meet these requirements.

It is well known that a linear inductor (L) and a capacitor (C) in series make an electromagnetic harmonic oscillator.
Wiring more and more inductors and capacitors together recursively into a periodic ladder gives the familiar transmission line along which electromagnetic waves can travel. Historically, various lattice structures of LC circuits have played an important role in network synthesis and filter design in the domain of electrical engineering \cite{kuo2006network}. They also serve as the lumped circuit models for continuum electromagnetic metamaterials \cite{caloz2005electromagnetic}. Thus most properties of LC circuits, including their periodic structures, have been extensively studied and well understood. Yet their topological aspects have rarely been discussed until very recently.

In this paper, we focus on certain periodic lattices of inductors and capacitors that are {\it wired differently}. At first sight, these circuits may appear utterly useless, or even nonsensical, from the traditional engineering perspective. But they possess one remarkable property: their frequency-wave number band structures, $\omega_m(\mathbf{k})$, are topologically nontrivial, reminiscent of electronic topological insulators and semimetals. Accordingly, via the bulk-boundary correspondence, there exist localized modes of electromagnetic excitations at the boundary or edge of the lattice. This begs a list of questions: what can serve as the building blocks of such topological circuits? What are the design principles? Can we use inductors and capacitors to study topological phenomena hard to reach in other classical many-body systems? What is the simplest topological LC circuit? And what is the natural language to describe their topological properties?

To address the first two questions, we proceed by following Nature's recipe for making crystal solids. First, we construct ``molecules" out of two kinds of ``atoms" (namely L and C) in the shape of loops, stars, and ladders as the building blocks. Then we design the coupling, or ``bonding",  between these ``molecules" by properly connecting them using L or C to form a periodic structure, i.e. a ``crystal."  Here a key concept is braided wiring between the building blocks which possess {internal degrees of freedom}, i.e., an emergent pseudospin structure. The ``weaving pattern" of L and C then dictates the frequency-wave number dispersion $\omega_m(\mathbf{k})$. We show that in this way, a surprisingly large variety of topological circuits can be built for {\it arbitrary effective spatial dimension} $D=1,2,3,4$ and beyond. This is an example of designing topological matter from bottom up, which in general is a highly nontrivial task \cite{bradlyn2017topological}.

Our work rests on a few key ideas put forward recently by other authors. Jia, Owens, Sommer, Schuster, and Simon designed and experimentally demonstrated the first topological LC circuit based on inductors with braided capacitive coupling \cite{PhysRevX.5.021031}. They realized the classic analog of two dimensional double Azbel-Hofstadter (dAH) model at flux $1/4$ (the Chicago design). Albert, Glazman and Jiang gave a general recipe to construct dAH model for any rational flux using capacitor loops with braided inductive coupling \cite{albert2015topological}. They gave a detailed discussion of how to achieve the dAH model at flux $1/3$ using the capacitor 3-loops (the Yale design) and also the symmetry protection of the edge modes. The insights from these authors will become pivotal in Section \ref{blocks}.

This paper goes beyond the aforementioned work in four aspects. (1) We discuss much simpler topological circuits, e.g. the one-dimensional SSH and braided ladder closely related to Creutz's ice-tray model, to illustrate the concepts of pseudospin, synthetic gauge fields, and symmetry protection in topological LC circuits. While the SSH circuit has been discussed previously \cite{PhysRevB.97.041106,lee-top-circ}, the perspective and approach here are different. (2) Different building blocks of topological circuits are proposed and treated within a single framework. In addition to capacitor loops, we also discuss inductor loops, capacitor stars, and inductor stars. The relationship between these designs, in particular the duality, is clarified. (3) We introduce and advocate a diagrammatic notation that drastically simplifies the presentation and design of topological LC circuits. For example, the Chicago and Yale design mentioned above appear clean and easier to understand in the new notation. (4) A list of new topological circuits are constructed. Using the simplified diagrammatic notation, we demonstrate how the circuit analogs of Dirac semimetal, Haldane's honeycomb lattice model of Chern insulators, and four-dimensional quantum Hall insulator (4D dAH model) can be realized.
We show that all these circuits can be understood using the same language and approach, which also establishes and clarifies the link between these ac circuits and quantum lattice models.

We emphasize that our main motivation is to show how a series of topological circuits could be designed following the idea pioneered in \cite{PhysRevX.5.021031} and \cite{albert2015topological}, and how a single theoretical framework can be used to understand their properties, including their connections to each other and to quantum lattice Hamiltonians describing topological insulators. We do not claim chronological originality in the design of any particular circuit. Many independent works have explored, for example, the SSH circuit \cite{PhysRevB.97.041106,lee-top-circ}, Dirac cone and higher dimensional circuits \cite{PhysRevB.97.041106}, Chern circuit \cite{hofmann2018chiral}, Weyl circuit \cite{lu2018probing} and topological corner modes \cite{ezawa2018higher,top-elec,ezawa2018electric}. Alternative theoretical approach can be found for example in \cite{PhysRevX.5.021031,helbig2018band}.

I hope the reader find it amusing to witness M\"obius strip, monopoles, geometric phases, and the first and second Chern numbers emerging from a tangled web of inducting coils, capacitor plates, and wires.

%To keep the discussion self-contained, we include definitions and algebras that may appear repetitive or straightforward to experts. Figures in this paper are self-explanatory and therefore not captioned.

\section{Lagrangian for circuits}

Consider an arbitrary lattice circuit consisting of linear inductors (L) and capacitors (C), i.e., coupled electromagnetic oscillators. It is conventional to describe the $j$-th element of a circuit using the current $i_j(t)$ through it and the voltage $v_j(t)$ across it. Alternatively, one can use the charge variable $q_j(t)$ and the flux variable $\phi_j(t)$. They are related to the current and voltage by
\be
q_j=\int i_j(t)dt,\;\;\; \phi_j=\int v_j(t)dt.
\ee
A dot denotes the time derivative, e.g., $v_j(t)=\dot \phi_j(t)\equiv d\phi_j(t)/dt$. Whenever possible, the time dependence is suppressed for brevity.

We describe linear circuits using Lagrangians. This approach is economical and convenient, because the analysis and calculation proceed pretty much the same way from one circuit to another. In addition, the language is familiar to physicists in the context of coupled oscillators or field theory. Consider a circuit described by $m$ charge variables, $\{q_j\}$ with $j=1,...,m$, and $n$ flux variables, $\{\phi_j\}$ with $j=m+1,...,m+n$. The action is given by
\be
S=\int dt \big( \mathscr{L}[q_j,\phi_j;\dot q_j,\dot \phi_j]+\sum_{j\leq m} q_j V_j + \sum_{j> m} \phi_j I_j \big).
\ee
Here $I_j$ and $V_j$ are external current and voltage sources, if any, coupled to the circuit.
From the action principle, $\delta S=0$, standard calculus of variations yields the Euler-Lagrange (EL) equation of motion
\begin{align}
V_j&=\frac{d}{dt}\frac{\partial \mathscr{L}} {\partial \dot{q}_j}-\frac{\partial \mathscr{L}} {\partial {q}_j},\\
I_j&=\frac{d}{dt}\frac{\partial \mathscr{L}} {\partial \dot{\phi}_j}-\frac{\partial \mathscr{L}} {\partial {\phi}_j}.
\end{align}
For circuits not connected to any external source, we simply set $V_j$ and $I_j$ to zero in the equations above.
Our notation follows that of Chua \cite{chua}.

The Lagrangian for a capacitor is $\La_C=-q^2/2C$, with $C$ the capacitance. Plugging it into the EL equation, we recover the defining relation for capacitor, $V=q/C$. Alternatively, we can take
\be
\La_C=\frac{C}{2}\dot{\phi}^2
\ee which gives $I=C\dot{V}$. These two descriptions are equivalent for ac circuits. For a linear inductor,
\be
\La_L=-\frac{1}{2L} \phi^2
\ee
where $L$ is the inductance. This gives via the EL equation $I=\phi/L$. We can also choose $\La_L=L\dot{q}^2/2$ which gives $V=L\ddot\phi =L\dot{I}$ as expected. In the representation based on the flux variable $\phi$, the mechanical analog of $\La_C$ is the kinetic energy, and $\La_L$ plays the role of potential energy \cite{albert2015topological}. The Lagrangian of the whole circuit is just the sum of the Lagrangian of all the capacitors and inductors.

\section{Connection and curvature}

The mathematical language describing the topological properties of periodic LC circuits is similar to that of electronic topological insulators and superconductors. But there are also key differences.
For a periodic circuit, the Lagrangian and the equations of motion are invariant under discrete translations of any lattice vector.
%The situation is rather similar to the phonon problem for lattice vibrations.
Via Bloch's theorem, the normal
modes of electromagnetic oscillations of the circuit can be labeled by the quasi-momentum $\mathbf{k}$  living in a $D$-dimensional Brillouin Zone (BZ), a $D$-torus $\mathbb{T}^{D}$. For all the circuits considered this paper, the equations of motion can be reduced to eigenvalue problems of the following form
\be
{H}(\mathbf{k})
u(\mathbf{k})=( a \omega^{b}+c)u(\mathbf{k}). \label{evp}
\ee
Here ${H}(\mathbf{k})$ is a Hermitian matrix which we call dynamical matrix or simply $H$-matrix, the Bloch wave function $u(\mathbf{k})$ is a column vector, $\omega$ is the frequency of oscillation,
$a$ and $c$ are two constants, and $b=\pm 2$. We stress that Eq. \eqref{evp} does not hold ture for arbitrary LC circuits.
Solving this eigenvalue problem yields the band structure $\omega_m(\mathbf{k})$, where the index $m$ labels the bands.
The $H$-matrix $H(\mathbf{k})$ plays a central role in our analysis. By design, it may possess certain symmetries.

Equation \eqref{evp} resembles the eigenvalue problem of a quantum Hamiltonian $H$ with energy eigenvalue $E\equiv a \omega^{b}+c$.
To make the connection more apparent, let us adopt Dirac's notation for the eigenvector $u$,
\be
H(\mathbf{k})|u_m(\mathbf{k})\rangle = E_m |u_m(\mathbf{k})\rangle,
\ee
which is simply a rewriting of Eq. \eqref{evp} above. The phase choice of each eigenvector $|u_m(\mathbf{k})\rangle$ is arbitrary,
for example, $|u_m(\mathbf{k})\rangle$ and $e^{i\theta_\mathbf{k}}|u_m(\mathbf{k})\rangle$ are equivalent. Each $\mathbf{k}$ point within the base
manifold $\mathbb{T}^{D}$ is thus associated with a fiber, namely a $U(1)$ space spanned by $e^{i\theta_\mathbf{k}}$. Fiber bundles of this type are well 
studied in quantum mechanics, e.g. in the context of Berry's phase. As one moves from $\mathbf{k}$ to $\mathbf{k}+d\mathbf{k}$,
a connection 1-form $A$, known as the gauge potential, can be defined by following the parallel transport of $|u_m(\mathbf{k})\rangle$.
Assume the $m$-th band is separated from other bands, the Berry connection $A_m$ is the 1-form
\be
A_m(\mathbf{k}) = \langle u_m(\mathbf{k})| du_m(\mathbf{k})\rangle. \label{b-A}
\ee
Here $d$ denotes the exterior derivative, and we follow the notation of Ref. \cite{ryu2010topological}. The Berry curvature $F_m$ is defined as the 2-form
\be
F_m(\mathbf{k})=dA_m(\mathbf{k}). \label{b-F}
\ee
Topological invariants for the Bloch bundles are defined as integrals of $A$ or $F$, depending on the spatial dimension and symmetry. These formulae are well known and can be found for example in Ref. \cite{ryu2010topological}.
In the rest of the paper, we will apply the framework outlined above, Eqs. \eqref{evp} to \eqref{b-F}, to
various circuits. To see how these equations come about and to make sense of them, let us first examine two simple examples in one dimension.

\section{Topological circuits in one dimension}
The transmission line is perhaps the most familiar one-dimensional (1D) periodic circuit. It is 
an infinite LC ladder (see figure below) and supports a gapless propagating mode with frequency $\omega(k)=2 \sin (k/2)/\sqrt{LC}$, which is proportional to the wave number $k$ in the long wavelength limit $k\rightarrow 0$ (we take the lattice constant to be unit length). By tweaking the transmission line a little bit, we can make 1D periodic circuits with topologically interesting band structures. They are the simplest topological circuits.

\ctikzset{bipoles/length=.8cm}
\begin{center}
\begin{circuitikz}
\draw (0,2)
	to [L=$L$] (2,2)
	to [C=$C$] (2,0)
	to [short] (0,0);
\draw (2,2)
	to [L=$L$] (4,2)
	to [C=$C$] (4,0)
	to [short] (2,0);
\draw (4,2)
	to [L=$L$] (6,2)
	to [C=$C$] (6,0)
	to [short] (4,0);
\draw (6,2)
	to [short] (6.5,2);
\draw (6,0)
	to [short] (6.5,0);
\end{circuitikz}
\end{center}

\subsection{SSH Circuit}
Let us double the unit cell of the transmission line above by introducing two alternating values of the inductance, $L_1\neq L_2$. This will open a gap in the spectrum. The diagram below shows the $n$-th unit cell, which contains two dynamic variables $\phi_1(n)$ and $\phi_2(n)$.
\begin{center}
\begin{circuitikz}
\draw (0,2)
	to [L=$L_1$] (2,2)	node[label={above:$\phi_1$}] {}
	to [C=$C$] (2,0)
	to [short] (0,0);
\draw (2,2)
	to [L=$L_2$, *-*] (4,2) node[label={above:$\phi_2$}] {}
	to [C=$C$] (4,0)
	to [short] (2,0);
\draw (4,2)
	to [short] (4.5,2);
\draw (4,0)
	to [short] (4.5,0);
\end{circuitikz}
\end{center}
The Lagrangian for the infinite ladder then has the form
\begin{align}
\La=&\sum_n \frac{C}{2} [\dot{\phi}_1^2(n) +\dot{\phi}_2^2(n) ] -  \frac{1}{2L_1} [{\phi}_1(n) -\phi_2(n-1)]^2 \nonumber \\
&-  \frac{1}{2L_2} [{\phi}_1(n) -\phi_2(n)]^2.
\end{align}
The EL equations of motion are
\begin{align}
C \ddot \phi_1(n)=-L^{-1}\phi_1(n)+L_1^{-1}\phi_2(n-1)+L_2^{-1}\phi_2(n), \nonumber \\
C \ddot \phi_2(n)=-L^{-1}\phi_2(n)+L_2^{-1}\phi_1(n)+L_1^{-1}\phi_1(n+1),
\end{align}
with the shorthand notation $L^{-1}=L_1^{-1}+L_2^{-1}$.
The $t$ dependence of $\phi_{j}$ is suppressed for brevity.

{\bf a. Band structure.} Assume oscillating solution and go over to the wave number $k$ space by Fourier transform,
\be
\phi_j(n;t)=e^{-i\omega t}\frac{1}{\sqrt{N}}\sum_k e^{ik n}\varphi_j (k), \;\; j=1,2,
\label{FT}
\ee
where $N$ is the number of unit cells and $k\in[-\pi,\pi]$. Then the EL equations in $k$ space become
\begin{align}
-C \omega^2 \left[
\begin{array}{c}
  \varphi_1   \\
  \varphi_2
\end{array}
\right]
=
\left[
\begin{array}{cc}
  -L^{-1} &    L_1^{-1}e^{-ik}+L_2^{-1}  \\
   L_1^{-1}e^{ik}+L_2^{-1}  &     -L^{-1}
\end{array}
\right]
 \left[
\begin{array}{c}
  \varphi_1   \\
  \varphi_2
\end{array}
\right]
\end{align}
Let us define
\be
\omega^2_0=1/C\sqrt{L_1L_2},\;\;\;  \eta=\sqrt{L_1/L_2}.
\ee
Then the eigenvalue problem becomes
\begin{align}
{H}(k)
 \left[
\begin{array}{c}
  \varphi_1(k)   \\
  \varphi_2 (k)
\end{array}
\right]=
\left(\frac{ \omega^2}{\omega^2_0}-\eta-\eta^{-1}\right) \left[
\begin{array}{c}
  \varphi_1(k)   \\
  \varphi_2 (k)
\end{array}
\right],
\label{h2}
\end{align}
with
\begin{align}
{H}(k)=
\left [
\begin{array}{cc}
0 &    -\eta^{-1}e^{-ik}-\eta   \\
   -\eta^{-1}e^{ik}-\eta   &     0
\end{array}
\right].
\end{align}
We see Eq. \eqref{h2} indeed has the form of Eq. \eqref{evp} with $b=2$.
Let us label the two Bloch bands by subscript $\pm$,
\be
\omega_\pm^2(k)/\omega^2_0=\eta+\eta^{-1} \pm \sqrt{\eta^2+\eta^{-2}+2\cos k }.
\ee
The lower band $\omega_-(k)\propto k$ as $k\rightarrow 0$ and thus is the ``acoustic" branch, while the upper band, the ``optical" branch, is separated from the lower branch by a finite energy gap. The gap closes when $\eta=1$, i.e., $L_1=L_2$. The figure below shows the
dispersion for $\eta=0.75$.

\begin{center}
  \includegraphics[width=0.4\textwidth]{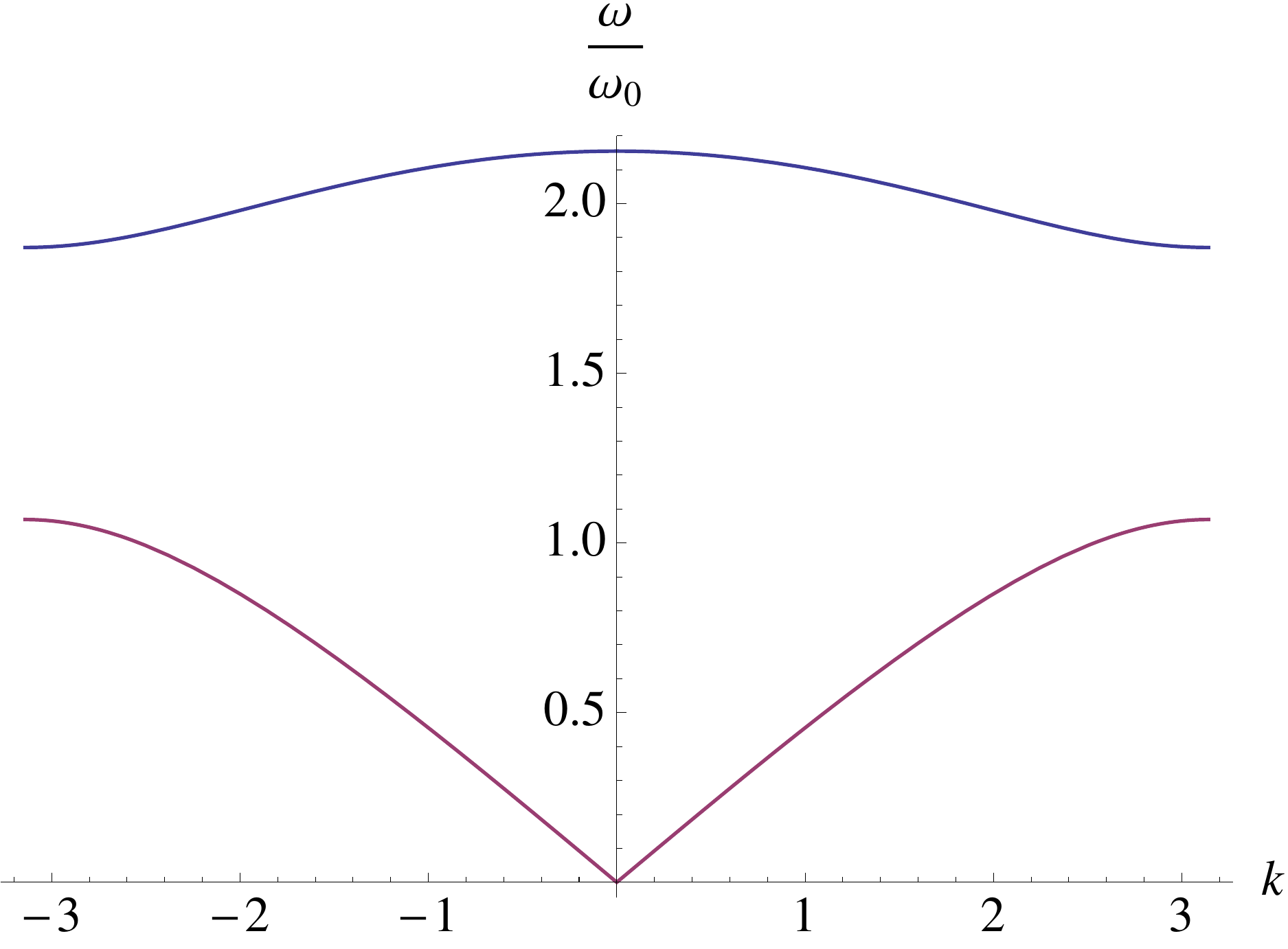}
\end{center}

{\bf b. H-matrix and winding number.}
The eigenvalue problem Eq. \eqref{h2} is analogous to the Su-Schrieffer-Heeger (SSH) model for spinless electrons \cite{su1979solitons}. The combination $\omega^2(k)/\omega^2_0-\eta-\eta^{-1}$ plays the role of the electron energy $E(k)$ while the dynamical matrix ${H}(k)$ corresponds to the Hamiltonian of the quantum lattice model. We can rewrite ${H}(k)$ in terms of Pauli matrices $\{\hat{\sigma}_i\}$,
\be
{H}(k)=\mathbf{d}(k)\cdot \boldsymbol{\sigma}=\sum_{i=x,y,z}  d_i(k){\sigma}_i.
\ee
The vector $\mathbf{d}(k)$ lies within the $xy$ plane,
\begin{align}
d_x&=-\eta-\eta^{-1}\cos {k},\\
d_y&=-\eta^{-1}\sin {k}, \\
d_z&=0.
\end{align}
As $k$ transverses the Brillouin Zone (which is topologically a circle $S^1$) from $-\pi$ to $\pi$, the unit vector $\hat{d}(k)=\mathbf{d}(k)/|\mathbf{d}(k)|$ winds by $2\pi$ for $\eta<1$, \\
\begin{center}
\includegraphics[width=0.4\textwidth]{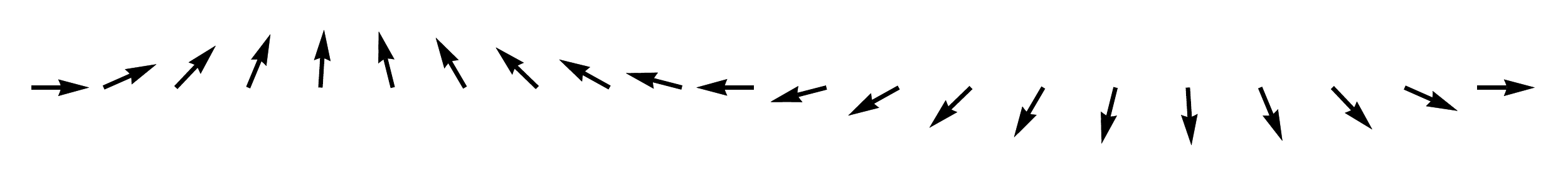}
\end{center}
In contrast, there is no net winding of $\hat{d}(k)$ for $\eta>1$, \\
\begin{center}
  \includegraphics[width=0.4\textwidth]{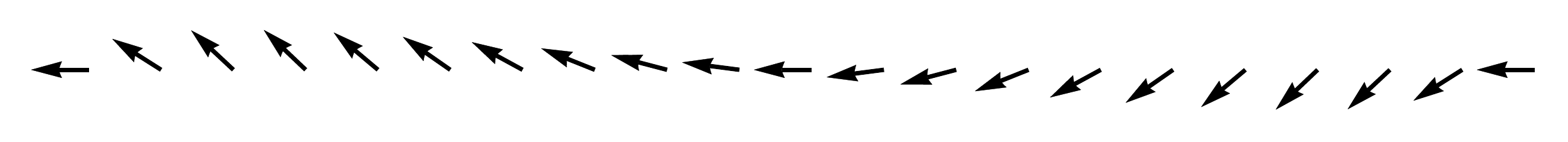}
\end{center}
The winding number $w$ can be defined through the turning of the arrows confined within the $xy$ plane,
\be
w = \frac{1}{2\pi}\int_{-\pi}^\pi dk[\hat{d}(k)\times \partial_k \hat{d}(k)]\cdot \hat{z}, \label{w-def1}
\ee
which gives $w=1$ for $\eta<1$, and  $w=0$ for $\eta>1$.
Thus ${H}(k)$ has two distinct topological sectors. It is impossible to deform an $H(k)$ with $w=1$ smoothly to $w=0$ without closing the gap.

{\bf c. Chiral symmetry.} The matrix ${H}(k)$ belongs to a class of Hermitian matrices that have sub-lattice or chiral symmetry: there exists a matrix ${\Gamma}$ that anti-commutes with $H$ and squares to one, $\Gamma H(k) \Gamma=-H(k)$ and $\Gamma^2=1$. Such chiral matrices can be brought into off-diagonal form in the eigen basis of $\Gamma$,
\be
{H}(k)={|\mathrm{Det} H |}
\left [
\begin{array}{cc}
0 &    q(k)   \\
  q^\dagger (k)   &     0
\end{array}
\right],
\ee
where $q$ is a $U(N)$ matrix in general. For the SSH circuit here, we have $\Gamma=\sigma_z$ and $q$ is just a complex number,
\be
q(k)=[d_x(k)-id_y(k)]/|\mathbf{d}(k)|\equiv e^{i\theta(k)}\in U(1). \label{q-ex}
\ee
Thus $q(k)$ defines a map from the $k$-space BZ, which is a circle $S^1$, to $U(1)$ which is also equivalent to $S^1$. The homotopy group of the mapping is $\pi_1(S^1)=Z$. And the topological invariant is the winding number
\be
w= \frac{i}{2\pi }\int_{BZ}q^{-1}dq=-\frac{1}{2\pi }\int_{-\pi}^\pi dk \partial_k \theta(k).
\ee
Evidently, this definition of $w$ agrees with the earlier definition Eq. \eqref{w-def1}.
Chiral symmetry is vital to have a meaningful definition of $q$ and $w$.

{\bf d. Berry connection and Zak phase.} We can define the winding number in yet another way using the
language of geometric phase, outlined in Section III. In Dirac notation, let
$| u_\pm(k)\rangle$ be the eigenvectors of $H(k)$,
\be
H(k)|u_\pm (k)\rangle = \pm  \sqrt{\eta^2+\eta^{-2}+2\cos k }  |u_\pm (k)\rangle.
\ee
Explicitly, $| u_\pm(k)\rangle$ stands for the column vector
\be
 |u_\pm (k)\rangle = \frac{1}{\sqrt{2}}\left [
\begin{array}{c}
q(k)   \\
 \pm 1
\end{array}
\right]. \label{u-ex}
\ee
And its conjugate transpose gives $\langle u_\pm(k)|$.
We can compute the Berry connection ${A}$ defined as
\be
{A}(k)= \langle u_+(k)|d u_+(k)\rangle. \label{berry}
\ee
From Eqs. \eqref{u-ex} and \eqref{q-ex}, we find
\be
{A}=\frac{i}{2} d\theta. \label{A2}
\ee
The same result is obtained if $A$ is defined using $u_-(k)$.
The curious factor of $1/2$ in Eq. \eqref{A2} has an interesting consequence.
As $k$ transverses the BZ to complete one circle, the state may not go back to itself,
just like the trip of an ant on the M\"obius strip.
%
%Loosely speaking, the pair $(k,\theta)$ defines a U(1)-bundle over
%the base space $S^1$ for $k$, where each fiber, the phase of $q(k)$, can also be visualized as a circle.
%
This phenomenon (holonomy) is described by the
geometric phase, also known as
the Zak phase in 1D periodic structures \cite{zak1989berry},
\be
\theta_Z \equiv i \int_{BZ} {A}=w\pi.
\ee
We see that the Zak phase is quantized: it is either $0$ or $\pi$ and proportional to the winding number.

%There is gauge redundancy in the definition of $H(k)$ and $q(k)$. One is free to choose the overall phase of the eigenvectors of $H(k)$, which amounts to a $U(1)$ transformation of $q(k)$.

{\bf e. Edge states.} For the topologically nontrivial case of $w=1$ and $\theta_Z=\pi$, edge states form within the bulk band gap
at frequency $\omega=\omega_0 \sqrt{\eta +\eta^{-1}}$. An example of the frequency spectrum of a finite  chain (40 unit cells) is shown below for $\eta=0.5$.
\begin{center}
\includegraphics[width=0.4\textwidth]{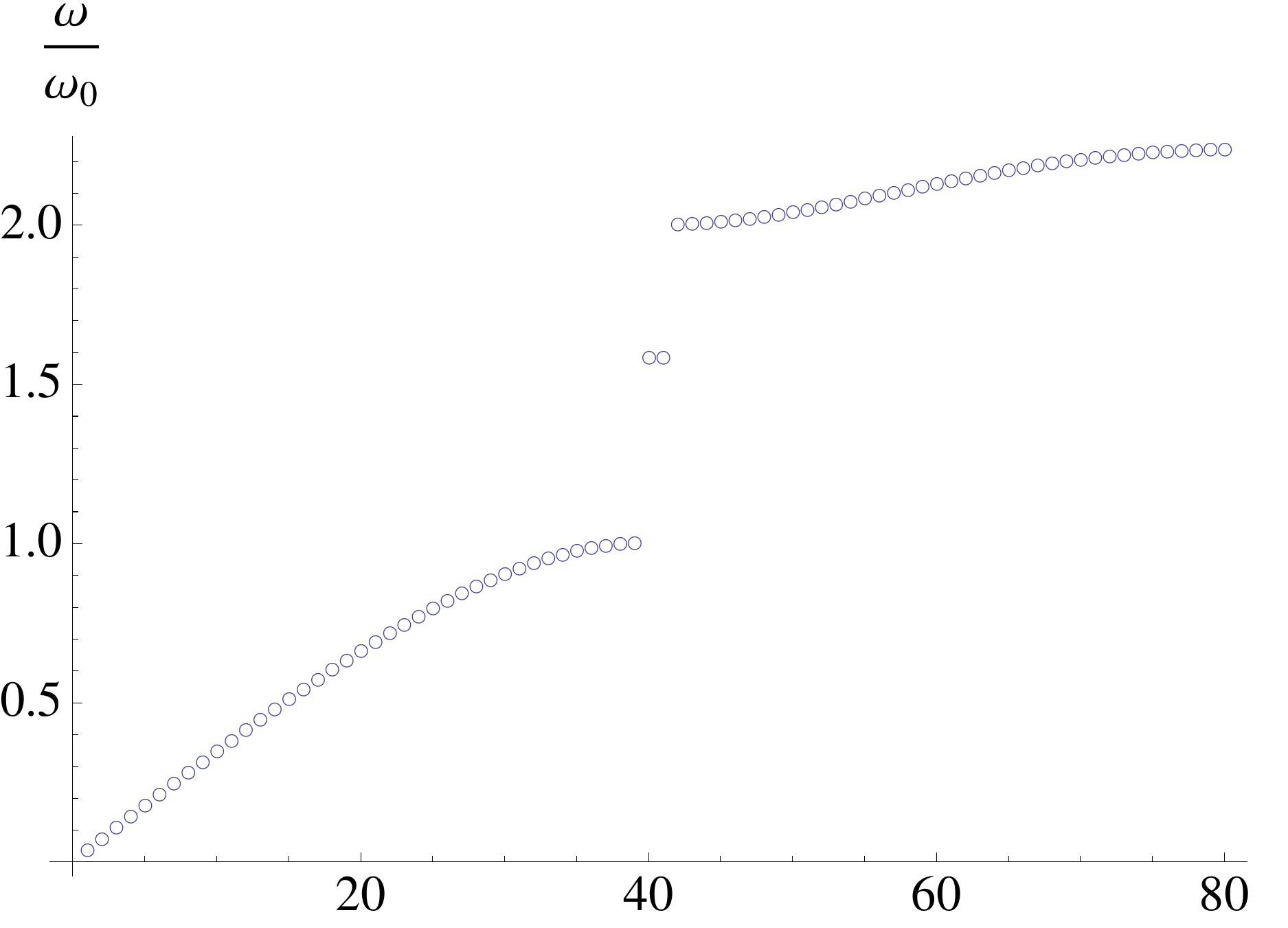}
\end{center}
In this case, the open end of the finite chain terminates with $L_2$, i.e. the inductor with larger inductance.
The decay of the edge state wave function is shown below.
% DO NOT DELETE!!!!
%
%\begin{center}
%\begin{circuitikz}
%\draw (2,2)
%	to [C=$C$] (2,0);
%\draw (2,2) node[label={above:$\phi_1$}] {}
%	to [L=$L_2$, *-*] (4,2) node[label={above:$\phi_2$}] {}
%	to [C=$C$] (4,0)
%	to [short] (2,0);
%\draw (4,2)
%	to [short] (4.5,2) ;
%\draw (4,0)
%	to [short] (4.5,0);
%\end{circuitikz}
%\end{center}
\begin{center}
\includegraphics[width=0.4\textwidth]{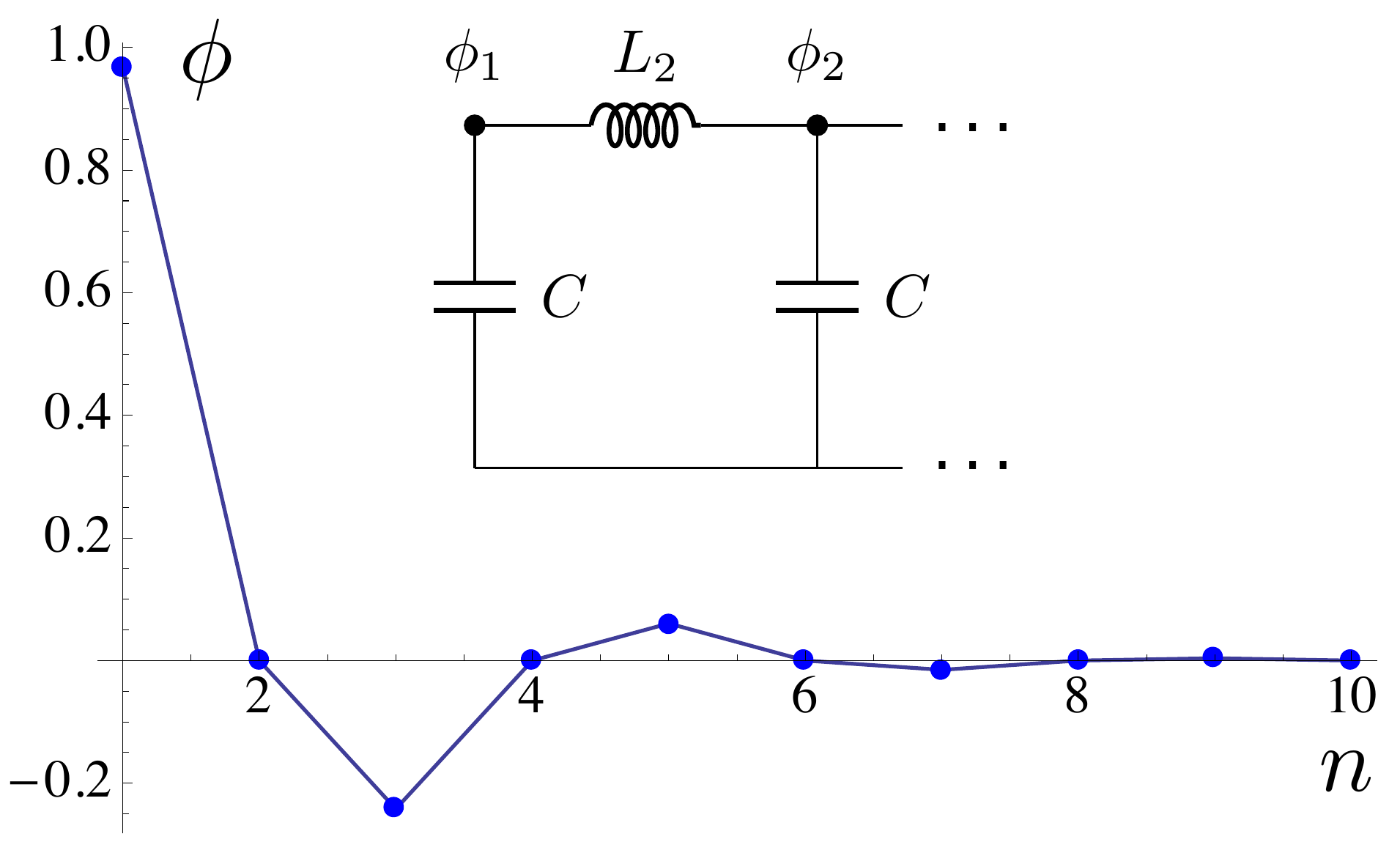}
\end{center}

% In contrast, for quantum many-body systems of fermions, time reversal symmetry and Kramers degeneracy plays an important role.
%One must, however, keep in mind that Fermi statistics or Kramers degeneracy does not apply to classical systems.

\subsection{Braided Ladder}

In our second example, we choose a symmetric layout for inductors along the two legs of the ladder, and the unit cell now contains a twist in the inductive wiring:

\ctikzset{label/align = straight}
\begin{center}
\begin{circuitikz}
\draw (-0.5,2)
	to [short] (0,2) node[label={above:$\phi^a_n$}] {}
	to [C,l_=$C$] (0,0)
	;
\draw (-0.5,0)
	to [short] (0,0) node[label={below:$\phi^b_n$}] {}
	;
\draw (0,2)
	to [L=$L_1$,*-*] (2,2) node[label={above:$\phi^c_n$}] {}
	to [C,l_=$C$] (2,0) node[label={below:$\phi^d_n$}] {}
	to [L=$L_1$,*-*] (0,0);
\draw (2,2)
	to [L=$L_2$] (3,1)
	to [short] (4,0)
	to [C,l_=$C$,*-*] (4,2)
	to [short] (3,1)
	to [L=$L_2$] (2,0);
\draw (4,2) 	node[label={above:$\phi^a_{n+1}$}] {}
	to [short] (4.5,2);
\draw (4,0) node[label={below:$\phi^b_{n+1}$}] {}
	to [short] (4.5,0);
\end{circuitikz}
\end{center}

\noindent The Lagrangian has the form
\begin{align}
\La &=\frac{C}{2} \sum_n [(\dot{\phi}^a_n-\dot{\phi}^b_n)^2 +(\dot{\phi}^c_n-\dot{\phi}^d_n)^2] \nonumber \\
&-  \frac{1}{2L_1} \sum_n[
(\phi^a_n -\phi^c_n)^2  + (\phi^b_n -\phi^d_n)^2 ] \nonumber \\
&- \frac{1}{2L_2}
\sum_n[(\phi^c_n -\phi^b_{n+1})^2  + (\phi^d_n -\phi^a_{n+1})^2].
\end{align}
With the ansatz
\be
\phi^\mu_n(t)=e^{-i\omega t}\frac{1}{\sqrt{N}}\sum_k e^{ik n}\varphi_\mu (k), \;\; \mu=a,b,c,d,
\ee
the EL equations become the eigenvalue problem
\begin{align}
&-C \omega^2
\left[
\begin{array}{rrrr}
 1 &   -1 & 0 & 0  \\
   -1 & 1 & 0 & 0 \\
0 & 0 & 1 & -1 \\
0 & 0 & -1 & 1
\end{array}
\right]
 \left[
\begin{array}{c}
  \varphi_a   \\
  \varphi_b \\
  \varphi_c   \\
  \varphi_d
\end{array}
\right] \label{4E} \\
&=
\left[
\begin{array}{llll}
  -L^{-1} & 0 & L_1^{-1} &    L_2^{-1}e^{-ik}  \\
  0 &   -L^{-1} & L_2^{-1}e^{-ik} & L_1^{-1} \\
  L_1^{-1} &    L_2^{-1}e^{ik} & -L^{-1} & 0 \\
  L_2^{-1}e^{ik} & L_1^{-1}  & 0 &  -L^{-1}
\end{array}
\right]
\left[
\begin{array}{c}
  \varphi_a   \\
  \varphi_b \\
  \varphi_c  \\
  \varphi_d
\end{array}
\right].\nonumber
\end{align}
As before, the shorthand notation $L^{-1}= L_1^{-1}+L_2^{-1}$.

We further diagonalize the matrix on the left hand side of Eq. \eqref{4E} by a
similarity transformation
\be
\left[
\begin{array}{cc}
  \varphi_a \\
   \varphi_b
\end{array}
\right]={U}\left[
\begin{array}{cc}
  \xi_1  \\
 \chi_1
\end{array}
\right],\;\;\;\left[
\begin{array}{cc}
  \varphi_c  \\
   \varphi_d
\end{array}
\right]={U}\left[
\begin{array}{cc}
  \xi_2  \\
 \chi_2
\end{array}
\right],
\ee
with
\be
{U}=\frac{1}{\sqrt{2}}
\left[
\begin{array}{rr}
 -1 &   1  \\
   1 & 1
\end{array}
\right].
\ee
Then the four equations in Eq. \eqref{4E} decouple into two sets.
The first set involves the flux difference (or voltage difference, after taking the time derivative)
$\xi_{1}=(\varphi_b-\varphi_a)/\sqrt{2}$ and $\xi_{2}=(\varphi_d-\varphi_c)/\sqrt{2}$,
\be
{H}(k)
 \left[
\begin{array}{c}
  \xi_1(k)   \\
  \xi_2 (k)
\end{array}
\right]=
\left(2\frac{ \omega^2}{\omega^2_0}-\eta-\eta^{-1} \right) \left[
\begin{array}{c}
  \xi_1(k)   \\
  \xi_2 (k)
\end{array}
\right],
\ee
where the $H$-matrix has chiral symmetry,
\be
{H}(k)={\sigma}_x(\eta \cos k - \eta^{-1})+{\sigma}_y \eta \sin k.
\label{hm2}
\ee
The second set involves flux average $\chi_{1}=(  \varphi_a+  \varphi_b)/\sqrt{2}$ and $\chi_{2}=(  \varphi_c+  \varphi_d)/\sqrt{2}$,
\begin{align}
0&=(L_1^{-1}+L_2^{-1}e^{-ik})\chi_2-L^{-1}\chi_1,\\
0&=(L_1^{-1}+L_2^{-1}e^{+ik}) \chi_1 -L^{-1}\chi_2.
\end{align}
For $\chi_{1,2}$, there is no effective capacitive coupling and consequently no oscillation.
A nonzero solution only exists for $k=0$, which gives constant $\chi_1=\chi_2$ throughout the circuit.
The similarity transform properly decouples the interesting dynamical degrees of freedom $\xi_{1,2}$ from the rest, $\chi_{1,2}$.
Such decoupling procedures will be used repeatedly in later sections.

{\bf a. Band structure.}
The dynamics of the braided ladder now reduces to the eigenvalue problem Eq. \eqref{hm2} which is similar to the SSH ladder
and takes the form of Eq. \eqref{evp} with $b=2$.
The band structure of $H(k)$ consists of two bands separated by a nonzero gap as long as $\eta\neq 1$,
\be
2\omega_\pm^2(k)/\omega^2_0=\eta+\eta^{-1} \pm \sqrt{\eta^2+\eta^{-2}-2\cos k }.
\ee
The minus sign before the cosine term marks the key difference from the SSH case. For example, the lower band $\omega_-(k)=0$ touches zero at $k=\pm \pi$ rather than $k=0$. The Zak phase and winding number can be introduced the same way as the SSH case and will not be repeated here. We find that the winding number $w=1$ (0) for $\eta>1$ ($\eta<1$), and edge states form for $w=1$.

{\bf b. Mapping to the $\pi$-flux ladder.} To connect the braided ladder to a quantum lattice model, we perform yet another rotation \cite{Guo2016}
\be
 \left[
\begin{array}{c}
  \alpha (k)   \\
  \beta (k)
\end{array}
\right]= V \left[
\begin{array}{c}
  \xi_1(k)   \\
  \xi_2 (k)
\end{array}
\right],\;\;\;\;\;
V=\frac{1}{\sqrt{2}}\left[
\begin{array}{rr}
 1 &   i  \\
   -1 & i
\end{array}
\right].\label{v-rot}
\ee
In the new basis, the $H$-matrix becomes
\be
H'(k)=V^\dagger H(k) V={\sigma}_z(\eta^{-1}-\eta \cos k )+{\sigma}_x \eta \sin k.
\ee
Fourier transform to real space, e.g. $\alpha(k)\rightarrow \alpha_n$, we find that $H'$ describes
a hopping problem,
\begin{align}
E \alpha_n=&+\eta^{-1} \alpha_n -\frac{\eta}{2} [\alpha_{n+1}+\alpha_{n-1}]
+i\frac{\eta}{2} [\beta_{n-1}-\beta_{n+1}], \nonumber\\
E \beta_n=&-\eta^{-1} \beta_n +\frac{\eta}{2}[\beta_{n+1}+\beta_{n-1}]
+i\frac{\eta}{2} [\alpha_{n-1}-\alpha_{n+1}]. \label{ice-tray}
\end{align}
Here we may view $E=2{\omega^2}/{\omega^2_0}-\eta-\eta^{-1}$ as energy, $\alpha_n$
and $\beta_n$ as wave functions on the two legs of a ladder shown in the figure below, where lines with arrows
represent hopping and their associated
phase factors ($\pm1$, $\pm i$) are indicated.
\begin{center}
\begin{circuitikz}
\draw (-0.5,2)
	to [short] (0,2) node[label={above:$\alpha_{n-1}$}] {}
	to [short,*-*,i=$-1$] (2,2) node[label={above:$\alpha_{n}$}] {}
	;
\draw (4.5,2)
	to [short] (4,2) node[label={above:$\alpha_{n+1}$}] {}
	to [short,*-*,i_=$-1$] (2,2)
	;
\draw (-0.5,0)
	to [short] (0,0) node[label={below:$\beta_{n-1}$}] {}
	to [short,*-*,i_=$1$] (2,0) node[label={below:$\beta_{n}$}] {}
	;
\draw (4.5,0)
	to [short] (4,0)	node[label={below:$\beta_{n+1}$}] {}
	to [short,*-*,i=$1$] (2,0)
	;
\draw[dashed] (2,0)
	to [short,i_=$i$] (1,1)
	to [short] (0,2)
	;
\draw[dashed]  (2,0)
	to [short,i=$-i$] (3,1)
	to [short] (4,2);
\draw [dashed] (2,2)
	to [short,i=$i$] (1,1)
	to [short] (0,0)
	;
\draw [dashed] (0,0)
	to [short] (2,2)
	to [short,i_=$-i$] (3,1)
	to [short] (4,0);
\end{circuitikz}
\end{center}

The total phase accumulated for a closed loop (e.g. the parallelogram $\alpha_{n-1}\rightarrow \alpha_n\rightarrow \beta_{n+1}\rightarrow \beta_n\rightarrow \alpha_{n-1}$) is $-1$. The situation is exactly the same as Creutz's ice tray model \cite{creutz2001aspects} which describes spinless electrons hopping on a ladder with magnetic flux of $\pi$ (half the flux quantum) threading each loop. Creutz gave a very intuitive explanation for the existence of edge states in the ice tray model \cite{creutz2001aspects}. Complete destructive interference for waves propagating, say, from $\alpha_{n-1}$ to $\beta_{n+1}$, along the two alternative paths leads to lack of diffusion, and consequently, edge modes localized at the boundary of the ladder.

{\bf c. Synthetic gauge field.} The braided ladder illustrates the concepts of ``synthetic gauge field" and ``synthetic dimension" which have been discussed in the context of cold atoms in optical lattice. Physically, $\alpha_n$ and $\beta_n$ are two independent modes that span the vector space of the internal degrees of freedom within each unit cell $n$. We may refer to them loosely as two different polarization states [see Eq. \eqref{v-rot}], or pseudo-spin up and down states, constructed from the linear combination of $\xi_{1,2}$. Alternatively we can visualize them as two ``sites" extending in a ``synthetic dimension" to form a rung, perpendicular to the physical spatial dimension labelled by unit cell index $n$. Moreover, braided wiring gives rise to a synthetic gauge field, an effective $\pi$-flux in the ladder system. The end result is the hopping model Eq. \eqref{ice-tray}.

{\bf d. Time reversal symmetry.} The Lagrangian for the SSH circuit or the braided ladder is quadratic in $\dot{\phi}_n$ and invariant under time reversal (TR). Accordingly, the equations of motion obey TR symmetry. Only $\omega^2$ appear in the
eigenvalue problem for $H$, and the band structure is degenerate for $\omega(k)$ and $-\omega(k)$. This is common for coupled classical oscillators where dissipation is negligible. Compared to many-body problem of fermions with Kramers degeneracy, the consequence of TR symmetry here is trivial.

%\subsection{Lesson from 1D}
For the simple 1D circuits above, the Lagrangian approach may seem an overkill. These circuits can be treated equally well by
conventional methods of network analysis, e.g. by computing the impedance or scattering matrix. Yet,
the Lagrangian approach provides a general route $\mathscr{L}\rightarrow H(k) \rightarrow \omega(k)$ that works for
 much more complicated circuits. Once the interesting degrees of freedom are identified  and described by $H$, it is
straightforward to map the problem to quantum lattice Hamiltonians or analyze the
topological invariants directly. The concepts of pseudospin, synthetic gauge field, and Berry connection
will be generalized to construct and analyze circuits in higher dimensions. In the next section, we will see that the braided ladder
is only a special case of a general construction scheme.

\section{Building blocks for topological circuits}\label{blocks}
To build more general topological circuits, we first construct various ``molecules" from $L$ and $C$ and study how they are coupled to each other. Obviously, there are infinite number of ways to hook them up. We will focus on the simplest constructions of loops and stars which enjoy a high degree of symmetry. One class of braided wiring, called ``shift," will play a key role in engineering the phase factors (synthetic gauge fields) in the $H$-matrix. Their mathematics turns out to be very neat. Our overall goal here is to have a {\it modular design}: these ``lego blocks" will be pieced together to form topological circuits in later sections. The blocks and wires will be {\it encapsulated}, i.e. with their inner details hidden in a simplified notation, to become nodes and $m$-connections respectively.

\subsection{Capacitor Loops}
Consider $p$ identical capacitors $C$ connected in series to form a loop labeled by $n$.
We will treat this case in great detail because all other designs are similar and can be easily understood by analogy.
The figure below shows a $p=4$ loop:
\begin{center}
\begin{circuitikz}
\draw (0,2) node[label={above:$\phi_0$}] {}
	to [C,*-*] (0,0) node[label={below:$\phi_1$}] {}
	to [C,*-*] (2,0) node[label={below:$\phi_2$}] {}
	to [C,*-*] (2,2) node[label={above:$\phi_3$}] {}
	to [C] (0,2)
	;
\end{circuitikz}
\end{center}
In terms of the flux variables $\{\phi_j(n)\}$, $j=0,..., p-1$, the Lagrangian of the loop
\be
\La_n \equiv  \frac{C}{2}\sum_{i,j=0}^{p-1} t_{ij}\dot{\phi}_i (n) \dot{\phi}_{j}(n) =\frac{C}{2}\sum_{j=0}^{p-1} [\dot{\phi}_j(n)-\dot{\phi}_{j+1}(n)]^2
\ee
with $\phi_p$ identified with $\phi_0$. Hence the capacitance matrix $t_{ij}$ is tridiagonal, resembling a tight binding
Hamiltonian describing particles hopping on a ring with onsite energy $t_{ii}=2$ and hopping amplitude $t_{i,i\pm 1}=-1$ between nearest neighbor sites, which is easily diagonalized by going to the quasi-momentum space. It is thus convenient to transform $\phi_j$ to $\varphi_k$,
\be
\phi_j(n)=\sum_k U_{jk}\varphi_k(n)=\frac{1}{\sqrt{p}}\sum_{k=0}^{p-1} e^{ i 2\pi k j/p}\varphi_k(n),
\label{phi-to-varphi}
\ee
which makes $\La_n$ diagonal,
\be
\La_n= 2C \sum_k [1-\cos ({2\pi}\frac{k}{p})] \varphi^2_k(n).
\ee
We can think of the capacitor loop as a benzene-like molecule, and interpret $k$ as its {\it eigenmode index}.

Now let us connect two such capacitor loops, labelled by $n$ and $n+1$ respectively, using $p$ identical inductors $L$. Each inductor connects a node ${\phi}_i (n)$ of the first loop to some node ${\phi}_{j}(n+1)$ of the second loop. The inductive coupling thus has the general form
\be
\La_{n,n+1} = -\frac{1}{2L}\sum_{i,j} v_{ij} {\phi}_i (n) {\phi}_{j}(n+1).
\ee
The wiring pattern can be illustrated with less clutter by hiding the capacitors as well as the wires connecting them within each loop, for example,
\begin{center}
\begin{circuitikz}
\draw (0,3) node[label={left:$\phi_0(n)$}] {}
	to [L,*-*] (1.5,1) node[label={right:$\phi_2(n+1)$}] {}
	;
\draw (0,2) node[label={left:$\phi_1(n)$}] {}
	to [L,*-*] (1.5,3) node[label={right:$\phi_0(n+1)$}] {}
	;
\draw (0,1) node[label={left:$\phi_2(n)$}] {}
	to [L,*-*] (1.5,2) node[label={right:$\phi_1(n+1)$}] {}
	;
\draw (0,0) node[label={left:$\phi_3(n)$}] {}
	to [L,*-*] (1.5,0) node[label={right:$\phi_3(n+1)$}] {}
	;
\draw[dashed,red] (5,3)
	to [short,*-*] (6.5,1)
	;
\draw[dashed,red] (5,2)
	to [short,*-*] (6.5,3)
	;
\draw[dashed,red] (5,1)
	to [short,*-*] (6.5,2)
	;
\draw[dashed,red] (5,0)
	to [short,*-*] (6.5,0)
	;
\end{circuitikz}
\end{center}
Furthermore, if we always align $\phi_j$ in some fixed order, say, $\phi_0$ to $\phi_3$ from top down, it is then unnecessary to label the nodes or draw the inductors explicitly. As shown in the figure above, we can simply use a dashed line to indicate a connection via $L$.
There are $p!$ different inductive wiring patterns, each corresponding to an element of the permutation group
of order $p$.

Let us focus on permutations given by
the {\it shift operation} $j\rightarrow j+m$, where modulo $p$ is implied and $m=0, ..., p-1$. In other words, we
connect the $j$-th capacitor within the $n$-th loop to the $(j+m)$-th capacitor within the $n+1$ loop,
\be
\La^{(m)}_{n,n+1} = -\frac{1}{2L}\sum_j [{\phi}_j (n)-{\phi}_{j+m}(n+1)]^2.
\ee
We will also call these connections the $m$-shift or $m$-twist.
The figure below shows the example of $p=4$. From left to right are the connection $m=0,1,2$, and $3$.
\begin{center}
\begin{circuitikz}
\draw[dashed,red] (0,1.5)
	to [short,*-*] (1,1.5)
	;
\draw[dashed,red] (0,1)
	to [short,*-*] (1,1)
	;
\draw[dashed,red] (0,0.5)
	to [short,*-*] (1,0.5)
	;
\draw[dashed,red] (0,0)
	to [short,*-*] (1,0)
	;
\draw[dashed,red] (2,1.5)
	to [short,*-*] (3,1)
	;
\draw[dashed,red] (2,1)
	to [short,*-*] (3,0.5)
	;
\draw[dashed,red] (2,0.5)
	to [short,*-*] (3,0)
	;
\draw[dashed,red] (2,0)
	to [short,*-*] (3,1.5)
	;
\draw[dashed,red] (4,1.5)
	to [short,*-*] (5,0.5)
	;
\draw[dashed,red] (4,1)
	to [short,*-*] (5,0)
	;
\draw[dashed,red] (4,0.5)
	to [short,*-*] (5,1.5)
	;
\draw[dashed,red] (4,0)
	to [short,*-*] (5,1)
	;
\draw[dashed,red] (6,1.5)
	to [short,*-*] (7,0)
	;
\draw[dashed,red] (6,1)
	to [short,*-*] (7,1.5)
	;
\draw[dashed,red] (6,0.5)
	to [short,*-*] (7,1)
	;
\draw[dashed,red] (6,0)
	to [short,*-*] (7,0.5)
	;
\end{circuitikz}
\end{center}

Now comes a crucial step. Let us express $\La^{(m)}$ in terms of $\varphi_k$. We find $\La^{(m)}$ not only becomes diagonal in $k$ but also picks up an interesting phase factor,
\begin{align}
\La^{(m)}_{n,n+1} = \frac{1}{2L}\sum_k  [-{\varphi}^2_k(n)-{\varphi}^2_k(n+1) \nonumber \\
+{\varphi}_k(n) {\varphi}_{k}(n+1)  e^{im2\pi k/p}]. \label{Lnk}
\end{align}
This result should not come across as a complete surprise. The phase factor $e^{im2\pi k/p}$ is nothing but the representation of the $m$-shift operation in basis $\{\varphi_k\}$.

The total Lagrangian of the two loops coupled by inductive wiring of type $m$ is then
\be
\La^ {(m)}= \La_n+\La_{n+1} +\La^{(m)}_{n,n+1}.
\label{Lnn+1}
\ee
The EL equation takes a rather clean form
\be
-2\frac{\omega^2}{\omega^2_0} [1-\cos ({2\pi}\frac{k}{p})]\varphi_k(n)=-\varphi_k(n) + e^{im2\pi k/p} \varphi_k(n+1) \label{el4}
\ee
where $\omega_0^2=1/LC$ as before. Again, if we make an analogy to quantum lattice models, the
 last term on the right hand side corresponds to a hopping amplitude with nontrivial phase factor  $e^{im2\pi k/p}$,
 depending  on the wiring pattern index $m$ and the molecular eigenmode index $k$.
This result is central to the design of topological circuits.

Next we show explicitly how pseudospin degrees of freedom emerge from the $p$-tuple $\{\varphi_k\}$. The procedure is best illustrated by a few examples.

{\bf (a) Four-loop.} Consider the $p=4$ capacitor loop. First note that for mode $k=0$, the prefactor of the $\omega^2$ term on the left hand side of Eq. \eqref{el4} (i.e. the effective capacitance) vanishes, so there is no dynamics associated with $\varphi_{k=0}$, just like what we have encountered in the braided ladder. The mode $\varphi_{k=1}$ and $\varphi_{k=3}$ are degenerate, with $1-\cos ({2\pi}{k}/{p})=1$. We may call them the pseudo-spin up and down mode, $\varphi_\uparrow=\varphi_1$ and $\varphi_\downarrow=\varphi_3$. They obey the EL equation
\be
-2\big(\frac{\omega}{\omega_0}\big)^2
\left[
\begin{array}{c}
  \varphi_{\uparrow}(n)   \\
  \varphi_{\downarrow}(n)
\end{array}
\right]
=-\left[
\begin{array}{c}
  \varphi_{\uparrow}(n)   \\
  \varphi_{\downarrow}(n)
\end{array}
\right]+
\left[
\begin{array}{r}
  (+i)^m\varphi_{\uparrow}(n+1)   \\
  (-i)^m\varphi_{\downarrow}(n+1)
\end{array}
\right]. \label{spin2}
\ee
%The phase factor for wiring pattern $m$ is $e^{\pm im\pi/2}=(\pm i)^m$,
Thus for $\varphi_{\uparrow}$, the $m$-th wiring pattern, the $m$-shift, is uniquely characterized by its corresponding phase factor $i^m$ in Eq. \eqref{spin2}. In the wiring diagram, it is sufficient to depict each $p=4$ loop using an empty circle and each wiring pattern $m$ using a single solid line together with its phase factor. The four types of inductive coupling shown above ($m=0,1,2,3$) then simplify to the $1$ connection, $i$ connection, $-1$ connection, and $-i$ connection below.
\begin{center}
\begin{circuitikz}
\draw[red] (0,0)
	to [short,o-]  (0.5,0) node[above] {$1$}
	to [short,-o] (1,0)
	;
\draw[red] (2,0)
	to [short,o-]  (2.5,0) node[above] {$i$}
	to [short,-o] (3, 0)
	;
\draw[red] (4,0)
	to [short,o-]  (4.5,0) node[above] {$-1$}
	to [short,-o] (5,0)
	;
\draw[red] (6,0)
	to [short,o-]  (6.5,0) node[above] {$-i$}
	to [short,-o] (7,0)
	;
\end{circuitikz}
\end{center}
The remaining $\varphi_{k=2}$  mode has a different eigen frequency,
\be
-4\big(\frac{\omega}{\omega_0}\big)^2
  \varphi_{2}(n) =- \varphi_{2}(n)+ (-1)^m \varphi_{2}(n+1).
  \label{4C}
\ee
Therefore it is energetically separated from the spin up and down mode.

{\bf (b) Three-loop.}  The case of $p=3$ is similar and worked out in detail in Ref. \cite{albert2015topological}. The  $\varphi_{0}$ mode has no dynamics since it describes the flux average of all the nodes within the loop. The  $\varphi_{1}$ and  $\varphi_{2}$ modes are degenerate with phase factor $e^{\pm i{m2\pi}/{3}}$,
\be
-3\big(\frac{\omega}{\omega_0}\big)^2
\left[
\begin{array}{c}
  \varphi_{1}(n)   \\
  \varphi_{2}(n)
\end{array}
\right]
=-\left[
\begin{array}{c}
  \varphi_{1}(n)   \\
  \varphi_{2}(n)
\end{array}
\right]+
\left[
\begin{array}{r}
  e^{+i\frac{m2\pi}{3}}\varphi_{1}(n+1)   \\
  e^{-i\frac{m2\pi}{3}}\varphi_{2}(n+1)
\end{array}
\right]. \label{3loop}
\ee
We can refer to them as the spin up and down mode respectively. The $m=0,1,2$ inductive wiring
\begin{center}
\begin{circuitikz}
\draw[dashed,blue] (0,1)
	to [short,*-*] (1.5,1)
	;
\draw[dashed,blue] (0,0.5)
	to [short,*-*] (1.5,0.5)
	;
\draw[dashed,blue] (0,0)
	to [short,*-*] (1.5,0)
	;
\draw[dashed,blue] (2.5,1)
	to [short,*-*] (4,0.5)
	;
\draw[dashed,blue] (2.5,0.5)
	to [short,*-*] (4,0)
	;
\draw[dashed,blue] (2.5,0)
	to [short,*-*] (4,1)
	;
\draw[dashed,blue] (5,1)
	to [short,*-*] (6.5,0)
	;
\draw[dashed,blue] (5,0.5)
	to [short,*-*] (6.5,1)
	;
\draw[dashed,blue] (5,0)
	to [short,*-*] (6.5,0.5)
	;
\end{circuitikz}
\end{center}
can be simply denoted by a single line with its phase factor for the $\varphi_1$ mode:
\begin{center}
\begin{circuitikz}
\draw[blue] (0,0)
	to [short,o-]  (0.75,0) node[above] {$1$}
	to [short,-o] (1.5,0)
	;
\draw[blue] (2.5,0)
	to [short,o-]  (3.25,0) node[above] {$e^{i\frac{2\pi}{3}}$}
	to [short,-o] (4,0)
	;
\draw[blue] (5,0)
	to [short,o-]  (5.75,0) node[above] {$e^{-i\frac{2\pi}{3}}$}
	to [short,-o] (6.5,0)
	;
\end{circuitikz}
\end{center}

{\bf (c) Two-loop.}  The simplest case is $p=2$. The 2-loop is trivial, since two capacitors in parallel amounts to a capacitor $2C$.
In other words, 2-loop is just a single capacitor.
\begin{center}
\begin{circuitikz}
\draw (0,1)
	to [C,l_=$C$] (0,0)
	to [short,-*] (0.5,0) node[label={below:$\phi_1$}] {}
	to [short] (1,0)
	to [C,l_=$C$] (1,1)
	to [short,-*] (0.5,1) node[label={above:$\phi_0$}] {}
	to [short] (0,1);
\draw (3,1) node[label={above:$\phi_0$}] {}
	to [C,l_=2$C$,*-*] (3,0) node[label={below:$\phi_1$}] {}
	;
\end{circuitikz}
\end{center}
Only the relative mode $\varphi_1=(\phi_0-\phi_1)/\sqrt{2}$ has oscillations when inductively coupled,
\be
-4\big(\frac{\omega}{\omega_0}\big)^2
  \varphi_{1}(n) =- \varphi_{1}(n)+ (-1)^m \varphi_{1}(n+1),
  \label{2C}
\ee
while the ``center of mass" mode $\varphi_0=(\phi_0+\phi_1)/\sqrt{2}$ is static and trivial.
There are only two kinds of wiring, $m=0$ and $1$,
\begin{center}
\begin{circuitikz}
\draw[dashed,teal] (0,1)
	to [short,*-*] (1.5,1)
	;
\draw[dashed,teal] (0,0.5)
	to [short,*-*] (1.5,0.5)
	;
\draw[dashed,teal] (2.5,1)
	to [short,*-*] (4,0.5)
	;
\draw[dashed,teal] (2.5,0.5)
	to [short,*-*] (4,1)
	;
\end{circuitikz}
\end{center}
which, according to our notation scheme above, simplify to
\begin{center}
\begin{circuitikz}
\draw[teal] (0,0)
	to [short,o-]  (0.75,0) node[above] {$1$}
	to [short,-o] (1.5,0)
	;
\draw[teal] (2.5,0)
	to [short,o-]  (3.25,0) node[above] {$-1$}
	to [short,-o] (4,0)
	;
\end{circuitikz}
\end{center}
We now recognize that these are exactly the building blocks for our braided ladder discussed earlier.
%By comparing Eq. \eqref{4C} with Eq. \eqref{2C}, we also recognize that the $\varphi_2$ mode of the
%$p=4$ loop, $\varphi_2=(\phi_0-\phi_1+\phi_2-\phi_3)/2$, is the relative mode of a $p=2$ loop of two effective
%capacitors $C/2$, each made of two capacitors in series.

To summarize, by connecting capacitor loops using inductors in braided fashion, pairs (except for $p=2$) of oscillation modes, referred to as pseudospin up and down mode $\varphi_{\uparrow,\downarrow}$, emerge to acquire hopping phase factor $e^{\pm im{2\pi}/{p}}$ respectively. These phase factors are the solutions of $z^p=1$ evenly distributed on the unit circle of the complex plane. So in principle, we can engineer almost any hopping phase factor by choosing some proper $p$.
In this subsection, we have also simplified the notation for the inductive wiring patterns to a single line labelled by its ``engineered" phase factor for the spin up mode $\varphi_\uparrow$. This simplification is required for handling complicated topological circuits in later sections.

\subsection{Capacitor Stars}
Following the wisdom of the $\Delta$-Y transformation, we are led to a design based on star structures of capacitors. The simplest is a three-star:
\begin{center}
\begin{circuitikz}
\draw (0,0) node[left] {$\phi_1$}
	to [C,*-*] (0.866,0.5) node[below] {$\phi_c$}
	to [C,*-*] (0.866,1.5) node[above] {$\phi_0$}
	;
\draw (0.866,0.5)
	to [C,*-*] (1.732,0)   node[right] {$\phi_2$}
	;
\end{circuitikz}
\end{center}
The total current flowing into the central node is zero from current conservation, $\sum_{i=0}^2 I_i=0$. The total charge of all plates connected to the central node is constant. Assuming the capacitors are initially uncharged, we have $\sum_i(\dot{\phi}_i-\dot{\phi}_c)/C=0$, or $\dot{\phi}_c=\sum_i \dot{\phi}_i/3$. So the voltage at the central node equals the the average voltage of $\dot{\phi}_i$. The Lagrangian for a general $p$-star labeled by $n$ is
 \begin{align}
\La_n&=\frac{C}{2}\sum_{j=0}^{p-1} \left[\dot{\phi}_j(n)-\dot{\phi}_c(n)\right]^2 \\
&=\frac{C}{2} (1-{p}^{-1}) \sum_{j} \dot{\phi}^2_j(n)-
\frac{C}{2}p^{-1} \sum_{j\neq i}\dot{\phi}_i(n)\dot{\phi}_j(n). \nonumber
\end{align}
We find that the capacitance matrix of the three-star is identical to the 3-loop, after rescaling $C \rightarrow 3C$. Therefore, all the discussions for inductive wiring and phase factors proceed the same way as before and will not be repeated here. Higher order stars with $p>3$ are less interesting because they are $(p-1)$-fold degenerate.
%For example, the four-star is equivalent to a tetrahedron.

\subsection{Inductor Loops}
The inductor loops are the dual design of the capacitor loops. An example is the inductor 3-loop below.
\begin{center}
\begin{circuitikz}
\draw (0,0) node[left] {$\phi_1$}
	to [L,*-*] (2,0) node[right] {$\phi_2$}
	to [L,*-*] (1,1.732) node[above] {$\phi_0$}
	to [L,*-*] (0,0)
	;
\end{circuitikz}
\end{center}
The Lagrangian for an inductor $p$-loop is
\be
\La_n =-\frac{1}{2L}\sum_{j=0}^{p-1} [{\phi}_j(n)-{\phi}_{j+1}(n)]^2.
\ee
For two such loops, labelled by $n$ and $n+1$, the coupling by $p$ identical capacitors according to a wiring pattern with shift $m$ is given by
\be
\La^{(m)}_{n,n+1} = \frac{C}{2}\sum_j [\dot{\phi}_j (n)-\dot{\phi}_{j+m}(n+1)]^2.
\ee
In terms of $\varphi_k$, the EL equation gives
\be
-2\frac{\omega_0^2}{\omega^2} [1-\cos ({2\pi}\frac{k}{p})]\varphi_k(n)=-\varphi_k(n) + e^{im2\pi k/p} \varphi_k(n+1).
\ee
Thus we arrive at a result very similar to the capacitor loop, only with $\omega/\omega_0$ replaced by $\omega_0/\omega$.
In passing, we note that
the duality
\be
L\leftrightarrow C, \;\;\; \frac{\omega}{\omega_0}\leftrightarrow \frac{\omega_0}{\omega}
\ee
also applies to the SSH and braided ladder circuits discussed above.

\subsection{Inductor Stars}
For completeness, we also mention the inductors 3-star.
\begin{center}
\begin{circuitikz}
\draw (0,0) node[left] {$\phi_1$}
	to [L,*-*] (0.866,0.5) node[below] {$\phi_c$}
	to [L,*-*] (0.866,1.5) node[above] {$\phi_0$}
	;
\draw (0.866,0.5)
	to [L,*-*] (1.732,0)   node[right] {$\phi_2$}
	;
\end{circuitikz}
\end{center}
Current conservation requires $\sum_i ({\phi}_i-{\phi}_c)/L=0$, or ${\phi}_c=\sum_i {\phi}_i/3$. The Lagrangian
 \begin{align}
\La_n &=-\frac{1}{2L}\sum_{j} \left[ {\phi}_j(n)- {\phi}_c(n)\right]^2 \\
&=-\frac{1}{2L} (1-{p}^{-1}) \sum_{j} {\phi}^2_j(n)+
\frac{1}{2L}p^{-1} \sum_{j\neq i}{\phi}_i(n){\phi}_j(n). \nonumber
\end{align}
For $p=3$, the inductive coupling is identical to the inductor 3-loop after rescaling $3L\rightarrow L$. Therefore, one can choose either one to construct topological circuits.

\subsection{Other Permutations}
Let us briefly comment on permutation wiring other than the $m$-shift, using the capacitor loops as examples. Within the subspace spanned by the spin up and down modes, all permutations are $2\times 2 $ matrices and can be decomposed in terms of the Pauli matrices. For example, for $p=3$, permutations of two nodes only
\begin{center}
\begin{circuitikz}
\draw[dashed,blue] (0,1)
	to [short,*-*] (1.5,1)
	;
\draw[dashed,blue] (0,0.5)
	to [short,*-*] (1.5,0)
	;
\draw[dashed,blue] (0,0)
	to [short,*-*] (1.5,0.5)
	;
\draw[dashed,blue] (2.5,1)
	to [short,*-*] (4,0)
	;
\draw[dashed,blue] (2.5,0.5)
	to [short,*-*] (4,0.5)
	;
\draw[dashed,blue] (2.5,0)
	to [short,*-*] (4,1)
	;
\draw[dashed,blue] (5,1)
	to [short,*-*] (6.5,0.5)
	;
\draw[dashed,blue] (5,0.5)
	to [short,*-*] (6.5,1)
	;
\draw[dashed,blue] (5,0)
	to [short,*-*] (6.5,0)
	;
\end{circuitikz}
\end{center}
give rise to off-diagonal matrix elements that couple spin up to down, i.e. the spin operator
\begin{center}
\begin{circuitikz}
\draw[blue] (0,0)
	to [short,o-]  (0.75,0) node[above] {$\sigma_x$}
	to [short,-o] (1.5,0)
	;
\draw[blue] (2.5,0)
	to [short,o-]  (3.25,0) node[above] {$\frac{-\sigma_x-\sqrt{3}\sigma_y}{2}$}
	to [short,-o] (4,0)
	;
\draw[blue] (5,0)
	to [short,o-]  (5.75,0) node[above] {$\frac{-\sigma_x+\sqrt{3}\sigma_y}{2}$}
	to [short,-o] (6.5,0)
	;
\end{circuitikz}
\end{center}
As another example, for $p=4$, the permutations
\begin{center}
\begin{circuitikz}
\draw[dashed,red] (0,1.5)
	to [short,*-*] (1.5,1.5)
	;
\draw[dashed,red] (0,1)
	to [short,*-*] (1.5,0)
	;
\draw[dashed,red] (0,0.5)
	to [short,*-*] (1.5,0.5)
	;
\draw[dashed,red] (0,0)
	to [short,*-*] (1.5,1)
	;
\draw[dashed,red] (2.5,1.5)
	to [short,*-*] (4,1)
	;
\draw[dashed,red] (2.5,1)
	to [short,*-*] (4,1.5)
	;
\draw[dashed,red] (2.5,0.5)
	to [short,*-*] (4,0)
	;
\draw[dashed,red] (2.5,0)
	to [short,*-*] (4,0.5)
	;
\end{circuitikz}
\end{center}
correspond to Pauli matrix
\begin{center}
\begin{circuitikz}
\draw[red] (0,0)
	to [short,o-]  (0.75,0) node[above] {$\sigma_x$}
	to [short,-o] (1.5,0)
	;
\draw[red] (2.5,0)
	to [short,o-]  (3.25,0) node[above] {$\sigma_y$}
	to [short,-o] (4,0)
	;
\end{circuitikz}
\end{center}
Interesting circuits containing such ``spin-flip" wiring $\sigma_i$ will be discussed elsewhere.

\section{Two dimensional topological circuits}
Now that the building blocks have been abstracted into nodes ($p$-loops or stars) and connections ($m$-shifts with phase factor $e^{im2\pi/p}$), we are ready to design topological circuits using these modular ``lego blocks." In this section, we retrofit a few well known quantum lattice models in two dimensions to illustrate the construction procedure. Along the way, we also review the Chicago design and the Yale design that inspired our work and present their layouts using the simplified notation.

\subsection{Dirac Cones}
\label{dd}
Our first 2D circuit is a square lattice array of capacitors connected by inductors. Each node consists of a single capacitor, which we may think of as the $p=2$ loop in the general scheme. There are two kinds of inductive connections, the $1$ connection ($m=0$, straight inductive wires) and the $-1$ connection with phase shift $\pi$ ($m=1$, twisted inductive wires). The $1$ and $-1$ connection are arranged alternatively in the following form:
\begin{center}
\begin{circuitikz}
\draw [teal] (2,0) to [short] (2.5,0) ;
\draw [teal] (2,0) to [short] (2,-0.5) ;
\draw [teal]  (2,2) to [short] (2.5,2)   ;
\draw [teal]  (2,2) to [short] (2,2.5);
\draw [teal]  (0,2) to [short] (-0.5,2) ;
\draw [teal,dashed,thick] (0,2.5)
	to [short] (0,2)
	to [short] (0,1)	node[left] {$-1$}
	to [short] (0,0)
	to [short] (0,-0.5)
	;
\draw [teal] (-0.5,0)
	to [short] (0,0) node [label={[shift={(-0.3,-0.7)}]$A$}]  {}
	to [short,o-]  (1,0) node[above] {$1$}
	to [short,-o] (2,0) node [label={[shift={(+0.3,-0.7)}]$B$}]  {}
	to [short]  (2,1) node[right] {$1$}
	to [short,-o] (2,2)
	to [short] (1,2)	node[above] {$1$}
	to [short,-o] (0,2)
	;
\end{circuitikz}
\end{center}
The unit cell contains two inequivalent nodes/sites $A$ and $B$.
The diagram above depicts the actual wiring below (dangling wires connecting to the rest of the lattice are not shown to avoid clutter):
\begin{center}
\begin{circuitikz}
\draw (0,1)
	to [C,*-*] (1,0)
	to [L]  (4,0)
	to [C] (3,1)
	to [L] (0,1)
	to [L] (1,3)
	to [C] (0,4)
	to [L] (0.5,2)
	to [short] (1,0)
	;
\draw (0,4)
	to [L,*-*] (3,4)
	to [C] (4,3)
	to [L,*-*] (1,3)
	;
\draw (3,4)
	to [L,*-*] (3,1)
	;
\draw (4,3)
	to [L,*-*] (4,0)
	;
\end{circuitikz}
\end{center}
Our simplified notation thus hides the wiring details to highlight the ideas behind the design. One may recognize that the circuit here mimics  the  Azbel-Hofstadter (AH) model \cite{azbel1964energy,PhysRevB.14.2239} at flux 1/2, a tight-binding model describing charged particles hopping on the square lattice in a perpendicular magnetic field with the magnetic flux per plaquttee being half the flux quantum. This is also known as the $\pi$-flux, since the total Peierls phase picked up by the particle after hopping around the plaquette is $\pi$, so the overall phase factor is $e^{i\pi}=-1$.

As a shortcut to find the band structure, we can take Eq. \eqref{2C} as our starting point. Let $\varphi(m,n)$ be the flux difference across the capacitor $C$ at lattice site labelled by a pair of integer coordinates $(x,y)=(m,n)$, we have
\begin{align}
-2 \omega^2
  \varphi(m,n) &=- z\varphi (m,n)+\varphi(m-1,n)+\varphi(m+1,n)\nonumber  \\
  &+ (-1)^m[\varphi(m,n-1)+\varphi(m,n+1)]
\end{align}
where $z=4$ is the coordination number of the lattice, and $\omega$ is in units of $\omega_0=1/\sqrt{LC}$. One can check that this equation indeed follows from the Lagrangian of the whole lattice.
In $k$-space, it leads to the eigenvalue problem
\be
\omega^2(k)  \left[
\begin{array}{c}
  \varphi_A(k)   \\
  \varphi_B (k)
\end{array}
\right]
= (2 +\cos k_y\sigma_z -\cos k_x \sigma_x)
 \left[
\begin{array}{c}
  \varphi_A(k)   \\
  \varphi_B (k)
\end{array}
\right].
\ee
The frequency spectrum is gapless. The two bands
\be
\omega_\pm(k) =2\pm \sqrt{\cos^2k_x +\cos^2k_y}
\ee
touch each other at a pair of Dirac points at the Brillouin zone boundary, $\mathbf{K}_\pm=(k_x,k_y)=(\pi/2,\pm\pi/2)$. Near the Dirac point, the spectrum is linear and takes the form of Dirac cones.

\subsection{The Yale Design}
A more interesting 2D topological circuit was proposed by Albert, Glazman and Jiang using capacitor 3-loops and permuted inductive wires \cite{albert2015topological}. We will call it the Yale design. In our notation, it takes the form
\begin{center}
\begin{circuitikz}
\draw [blue] (0,2.5)
	to [short] (0,2)
	to [short,l_=$1$] (0,0)
	to [short] (0,-0.5)
	;
\draw [blue,dashed,thick] (2,2.5)
	to [short] (2,2)
	to [short,l_=$e^{i\frac{2\pi}{3}}$] (2,0)
	to [short] (2,-0.5)
	;
\draw [blue,dotted, thick] (4,2.5)
	to [short] (4,2)
	to [short,l_=$e^{-i\frac{2\pi}{3}}$] (4,0)
	to [short] (4,-0.5)
	;
\draw [blue] (-0.5,0)
	to [short] (0,0)
	to [short,l_=$1$,o-o]  (2,0)
	to [short,-o] (4,0)
	to [short]  (4.5,0)
	;
\draw [blue] (-0.5,2)
	to [short] (0,2)
	to [short,l^=$1$,o-o]  (2,2)
	to [short,-o] (4,2)
	to [short]  (4.5,2)
	;
\end{circuitikz}
\end{center}
Here each node is a capacitor 3-loop, and three kinds of inductive wirings with phase factor $e^{im2\pi/3}$ are used. The circuit corresponds to the double Azbel-Hofstadter (dAH) model at flux $1/3$. The spectrum has three bands separated by two gaps. And each band is characterized by a nonzero Chern number. This guarantees the existence of edge states within the band gaps.
The actual layout and more details can be found in Ref. \cite{albert2015topological} and will not be repeated here.
Note that a few variants of the Yale design can be created. As we have argued above, the node could be replaced by a capacitor 3-star, inductor 3-loop, or inductor 3-star. For each choice of the node, one needs to use the corresponding $m$-connection.

\subsection{The Chicago Design}
The first topological LC circuit was designed and experimentally demonstrated by Jia, Owens, Sommer, Schuster, and Simon \cite{PhysRevX.5.021031}. They introduced the ingenious idea of braided capacitor coupling. We will refer to it as the Chicago design. In the simplified notation, it has the following layout
\begin{center}
\begin{circuitikz}
\draw [teal] (0,1.25)
	to [short] (0,1)
	to [short,l_=$1$] (0,0)
	to [short] (0,-0.25)
	;
\draw [teal] (1,1.25)
	to [short] (1,1)
	to [short] (1,0)
	to [short] (1,-0.25)
	;
\draw [teal] (2.25,1.25)
	to [short] (2,1)
	to [short] (3,0)
	to [short] (2.75,-0.25);
\draw [teal,dashed,thick] (2.75,1.25)
	to [short] (3,1)
	to [short] (2,0)
	to [short] (2.25,-0.25);
\draw [teal,dashed,thick] (4,1.25)
	to [short] (4,1)
	to [short] (4,0)
	to [short] (4,-0.25)
	;
\draw [teal,dashed,thick] (5,1.25)
	to [short] (5,1)
	to [short,l_=$-1$] (5,0)
	to [short] (5,-0.25)
	;
\draw  [teal,dashed,thick] (6.25,1.25)
	to [short] (6,1)
	to [short] (7,0)
	to [short] (6.75,-0.25);
\draw [teal] (6.75,1.25)
	to [short] (7,1)
	to [short] (6,0)
	to [short] (6.25,-0.25);
\draw [teal] (-0.25,0)
	to [short, -o] (0,0)
	to [short, -o] (1,0)
	to [short, -o] (2,0)
	to [short, -o] (3,0)
	to [short, -o] (4,0)
	to [short, -o] (5,0)
	to [short, -o] (6,0)
	to [short, -o] (7,0)
	to [short] (7.25,0)
	;
\draw [teal] (-0.25,1)
	to [short, -o] (0,1)
	to [short, -o] (1,1)
	to [short, -o] (2,1)
	to [short, -o] (3,1)
	to [short, -o] (4,1)
	to [short, -o] (5,1)
	to [short, -o] (6,1)
	to [short, -o] (7,1)
	to [short] (7.25,1)
	;
\end{circuitikz}
\end{center}
Here each node is an inductor (i.e. a $p=2$ inductor loop in the general scheme). The dashed line is the braided $m=1$ capacitive connection with phase factor $-1$, while the solid line is the $m=0$ connection with phase factor $1$. The unit cell is quite large and consists of four different layers, each forming a straight or braided ladder that extends in the $y$-direction. The whole circuit  realizes the flux $1/4$ dAH model.

Alternatively, we can follow the idea of the Yale design to realize the flux $1/4$ dAH model by using the capacitor 4-loops as nodes and the connection pattern
\begin{center}
\begin{circuitikz}
\draw [red] (0,1.25)
	to [short] (0,1)
	to [short,l=$1$] (0,0)
	to [short] (0,-0.25)
	;
\draw [red,dotted,thick] (1,1.25)
	to [short] (1,1)
	to [short,l=$i$] (1,0)
	to [short] (1,-0.25)
	;
\draw [red,dashed, thick] (2,1.25)
	to [short] (2,1)
	to [short,l=$-1$] (2,0)
	to [short] (2,-0.25);
\draw [red,densely dotted, thick] (3,1.25)
	to [short] (3,1)
	to [short,l=$-i$] (3,0)
	to [short] (3,-0.25);

\draw [red] (-0.25,0)
	to [short, -o] (0,0)
	to [short, -o] (1,0)
	to [short, -o] (2,0)
	to [short, -o] (3,0)
	to [short] (3.25,0)
	;
\draw [red] (-0.25,1)
	to [short, -o] (0,1)
	to [short, -o] (1,1)
	to [short, -o] (2,1)
	to [short, -o] (3,1)
	to [short] (3.25,1)
	;
\end{circuitikz}
\end{center}
Compared to the Chicago design, the unit cell looks simpler. But each node (empty circle) now is a 4-capacitor loop, so the unit cell actually contains more circuit elements.

\subsection{Magnetic Monopole}
The flux $1/2$ AH circuit in Section VI.A has no spectral gap. To construct the analog of quantum Hall insulator at flux 1/2, we need to gap out the  Dirac spectrum. One mechanism to achieve this is by introducing coupling between the next nearest neighbors, i.e. adding more wires and inductors. In addition, we need to use capacitor 4-loops as nodes (the lattice sites), which offer four kinds of inductive connections (hopping between sites) with phase factor $\pm1$ and $\pm i$ respectively. This leads to the following circuit design.
\begin{center}
\begin{circuitikz}
\draw [red] (0,2.5)
	to [short] (0,1) node[left] {$1$}
	to [short] (0,0) node [label={[shift={(-0.3,-0.7)}]$A$}]  {}
	to [short] (0,-0.5)
	;
\draw [red,dashed,thick] (2,2.5)
	to [short] (2,1) node[left] {$-1$}
	to [short] (2,0) node [label={[shift={(0.3,-0.7)}]$B$}]  {}
	to [short] (2,-0.5)
	;
\draw [red] (4,2.5)
	to [short] (4,2)
	to [short] (4,0)
	to [short] (4,-0.5)
	;
\draw [red,dotted,thick] (0,2)
	to [short] (2,0);
\draw [red,dotted,thick] (0,0)
	to [short] (1.2,1.2) node[above] {$i$}
	to [short] (2,2)
	;
\draw [red,densely dotted] (2,2)
	to [short] (4,0)
	;
\draw [red,densely dotted] (2,0)
	to [short] (3.2,1.2) node[above] {$-i\;\;$}
	to [short] (4,2)
	;
\draw [red] (-0.5,0)
	to [short] (0,0)
	to [short,o-o] (2,0)
	to [short,-o]  (4,0)
	to [short] (4.5,0)
	;
\draw [red] (-0.5,2)
	to [short] (0,2)
	to [short,o-o]  (2,2)
	to [short,-o] (4,2)
	to [short]  (4.5,2)
	;
\end{circuitikz}
\end{center}
Each unit cell contains two nodes $A$ and $B$.
To work out the band structure, let us start from Eq. \eqref{spin2} and consider the pseudo spin up mode $\varphi_\uparrow$ for example.
In $k$ space, its eigenvalue problem takes the form
\be
\left(\omega^2(k)-\frac{1}{2}\right) \left[
\begin{array}{c}
  \varphi_{\uparrow A}(k)   \\
  \varphi_{\uparrow B} (k)
\end{array}
\right] = H(\mathbf{k})
\left[
\begin{array}{c}
  \varphi_{\uparrow A}(k)   \\
  \varphi_{\uparrow B} (k)
\end{array}
\right],
\ee
where the $H$-matrix
\be
H(\mathbf{k}) = - \mathbf{h}(\mathbf{k})\cdot \boldsymbol{\sigma}
\ee
with $\boldsymbol{\sigma}=(\sigma_x,\sigma_y,\sigma_z)$ the Pauli matrices
and $\mathbf{h}$ is a fictitious magnetic field
\be
\mathbf{h}=(h_x,h_y,h_z)=(\cos k_x, -2\sin k_x\sin k_y, \cos k_y). \label{expression-h}
\ee
The eigen frequency (in unit of $\omega_0$) is then given by
\be
\omega^2_\pm(k) =\frac{1}{2}\pm |\mathbf{h}(\mathbf{k})|. \label{sp-down}
\ee
The spectrum contains two bands labelled by $\pm$ and separated by a finite gap.
The eigenvalue problem is similar to the generalized flux 1/2 AH model considered by Hatsugai and Kohmoto \cite{PhysRevB.42.8282}.
The eigen equation for $\varphi_\downarrow$ is the complex conjugate of that
for $\varphi_\uparrow$. Namely, the spin up and down mode experience opposite synthetic
magnetic field, hence the ``double" in double AH model. The spectrum
of $\varphi_\downarrow$ is degenerate to $\varphi_\uparrow$ and also given by Eq. \eqref{sp-down},
a manifestation of time reversal symmetry.

The unit vector $\hat{h}(\mathbf{k})$ defines a mapping from the two dimensional BZ, a torus $T^2$, to a sphere $S^2$.
One can verify that as one exhausts the $\mathbf{k}$ points within the BZ, $0\leq k_x \leq\pi$ and $-\pi\leq k_y\ \leq \pi$, the corresponding
$\hat{h}(\mathbf{k})$ will cover the whole $S^2$ once. For example, $\mathbf{k}_S=(\pi/2,\pm\pi)$ maps to the south pole, and $\mathbf{k}_N=(\pi/2,0)$ to the north pole, while the two cuts $k_y=\pm \pi/2$ across the BZ correspond to the equator.
This mapping is characterized by an integer topological invariant $c$, since the homotopy group $\pi_2(S^2) = Z$.
The integer $c$ is known as the Pontryagin index
\be
c\equiv \frac{1}{4\pi}\int _{BZ} \frac{1}{\;|\mathbf{h}|^3} \mathbf{h}\cdot (\partial_x\mathbf{h}\times \partial_y\mathbf{h}) d k_x d k_y.
\label{pontry}
\ee
Plugging in the expression for $\mathbf{h}$ from Eq. \eqref{expression-h} and performing the integral, we find that $c=1$.
The number $c$ has a clear physical interpretation.
In the space of $\mathbf{h}$, the set of vectors $\{\hat{h}(\mathbf{k})\}$ span the unit sphere. Then $c$ is the total magnetic flux out of the sphere, i.e., the charge of magnetic monopole inside the sphere. In our circuit model, the monopole has unit charge. For some other circuits, $\{\hat{h}(\mathbf{k})\}$ may cover the unit sphere twice. Then $c=2$ and we will have a charge-2 monopole.

As before, the band topology can be described in the general language of bundles and connections.
%, so it applies to arbitrary $H$-matrices and BZ.
Introduce the eigenvectors in Dirac notation,
\be
H(\mathbf{k})|u_\pm(\mathbf{k})\rangle = E_\pm |u_\pm(\mathbf{k})\rangle,
\ee
and consider for example the lower band, $E_-=-|\mathbf{h}|$.
The Berry connection $A$ is the 1-form
\be
A = \langle u_- | du_-\rangle,
\ee
and the Berry curvature $F$ is the 2-form
\be
F=dA,
\ee
where $d$ denotes the exterior derivative.
%so that these expressions are independent of the choice of coordinates.
%At each $\mathbf{k}$ point, there is vector space associated with $H(\mathbf{k})$. vector bundle
The topological invariant is the first Chern number
\be
c=\frac{i}{2\pi}\int_{BZ} F. \label{chnum}
\ee
In our case, $H=- \mathbf{h}(\mathbf{k})\cdot \boldsymbol{\sigma}$, both $A$ and $F$ can be evaluated following
Berry's treatment of spin 1/2 problem \cite{berry1984quantal},
\be
F=-\frac{i}{4}|\mathbf{h}|^{-3}\epsilon^{ijk} h_i dh_j \wedge d h_k.
\ee
Plugging this expression into Eq. \eqref{chnum}, we recover Eq. \eqref{pontry}: the first Chern number coincides with the Pontryagin index introduced earlier. For the pseudo spin up mode, the Chern number for the lower (upper) band is $c=1$ ($c=-1$). The Chern number for the pseudo spin down mode is the opposite of the spin up mode.

It would be messy to draw all the capacitors and inductors explicitly: there are 32 wires coming out of each node.
Compared to the Chicago and Yale design, this circuit is not simpler in its actual layout. But as a two-band model, its mathematics is very elegant. Using only inductors and capacitors, we have engineered a magnetic monopole living in the parameter space of $\hat{h}(\mathbf{k})$.

\subsection{Haldane Circuit}

As the final example in 2D, we show that the building blocks are sufficient to construct the analog of Haldane's model of Chern insulators \cite{PhysRevLett.61.2015}. Let us arrange capacitor 3-loops (empty circles), inductive 1-connection (solid line), $e^{ i2\pi/3}$ connection (dashed line), and $e^{-i2\pi/3}$ connection (dotted line) into a honeycomb lattice:
\begin{center}
\begin{circuitikz}
\draw [dashed, blue,thick] (0,0)
	to [short]  (1.5,0.866)
	to [short,]  (0,1.732)
	to [short] (0,0) node [label={[shift={(-0.35,-0.3)}]$A$}]  {}
	;
\draw [dotted, blue, thick] (1,0)
	to [short] (1,1.732)
	to [short] (-0.5,0.866)
	to [short] (1,0) node [label={[shift={(0.3,-0.3)}]$B$}]  {}
	;
\draw [blue,thick] (0,0)
	to [short,o-o] (1,0)
	to [short,-o]  (1.5,0.866)
	to [short,-o] (1,1.732)
	to [short,-o]  (0,1.732)
	to [short,-o] (-0.5,0.866)
	to [short] (0,0)
	;
\end{circuitikz}
\end{center}
The inductors used in the solid (dash/dotted) connection have inductance $L_2$ ($L_1$), and in general $L_1\neq L_2$.
It is easier to analyze this circuit directly from Eq. \eqref{3loop},
but in order to illustrate the calculation procedure, we will start from
 the Lagrangian for the whole lattice,
\begin{align}
\mathscr{L}=&\frac{C}{2}\sum_{\v{r},j}\left[(\dot{\phi}_{jA}(\v{r})-\dot{\phi}_{j+1,A}(\v{r}))^2+(A\rightarrow B)\right] \nonumber \\
-&\frac{1}{2L_1}\sum_{\v{r},j,\gamma} \left(\phi_{jA}(\v{r})- \phi_{j+1,A}(\v{r}+\v{b}_\gamma)\right)^2 \nonumber  \\
-&\frac{1}{2L_1}\sum_{\v{r},j,\gamma} \left(\phi_{jB}(\v{r})- \phi_{j-1,B}(\v{r}+\v{b}_\gamma)\right)^2 \nonumber \\
-&\frac{1}{2L_2}\sum_{\v{r},j,\gamma} \left(\phi_{jA}(\v{r})- \phi_{j,B}(\v{r}+\v{a}_\gamma)\right)^2.
\end{align}
Here $\v{r}$ is the lattice vector labeling the unit cell (with two sites $A$ and $B$), $j=0,1,2$ labels the flux variables inside each capacitor 3-loop,
$\v{a}_\gamma$ and $\v{b}_\gamma$ are vectors connecting the nearest neighbor and the next nearest neighbor respectively:
$\v{a}_1=(1,0)$, $\v{a}_1=(-1/2,\sqrt{3}/2)$, $\v{a}_3=-\v{a}_1-\v{a}_2$; $\v{b}_1=\v{a}_2-\v{a}_3$, $\v{b}_2=\v{a}_3-\v{a}_1$,
$\v{b}_3=\v{a}_1-\v{a}_2$.
Carry out the differentiation to find the EL equations of motion and then Fourier transform
\be
\phi_{jA}(\v{r})=\frac{1}{\sqrt{3N}}\sum_{\v{k},\ell}\varphi_{\ell A}(\v{k})e^{i\v{k}\cdot \v{r}+i2\pi\ell j/3}.
\ee
We find the equations for different modes $\ell$ decouple. For example, for the $\ell=1$ (spin up) mode, we once again
have a $2\times 2$ eigenvalue problem,
\be
\left(3\omega^2(\v{k})-9\right) \left[
\begin{array}{c}
  \varphi_{\ell A}(\v{k})   \\
  \varphi_{\ell B} (\v{k})
\end{array}
\right] = H(\mathbf{k})
\left[
\begin{array}{c}
  \varphi_{\ell A}(\v{k})   \\
  \varphi_{\ell B} (\v{k})
\end{array}
\right],
\ee
with an $H$-matrix
\be
H(\mathbf{k})=
\left[
\begin{array}{cc}
\eta \beta^+_\v{k} &  \alpha_\v{k}  \\
 \alpha^*_\v{k} &  \eta \beta^-_\v{k}
\end{array}
\right].
\ee
As before, $\omega$ is measured in unit of $\omega_0=1/\sqrt{L_1C}$, and $\eta=L_1/L_2$. The
form factors are defined by
\be
\alpha_\v{k} = \sum_{\gamma}e^{i\v{k}\cdot \v{a}_\gamma},\;\;
\beta^\pm_\v{k} = \sum_{\gamma}e^{i(\v{k}\cdot \v{b}_\gamma\pm 2\pi/3)} + h.c. \label{formfactors}
\ee
Comparing to Haldane's model for spinless electrons \cite{PhysRevLett.61.2015}, we see that $E=3\omega^2(\v{k})-9$
plays the role of electron energy, $\eta$ is the hopping ratio $t'/t$, the staggered potential is set to zero, and the flux $\Phi=2\pi/3$.
The equations for the spin down mode, $\ell=2$, take the same form except for a replacement $2\pi/3\rightarrow -2\pi/3$
in Eq. \eqref{formfactors}. So we arrive at a double copy of Haldane's model, where the effective magnetic flux for two pseudospins is opposite of each other.

%The properties of Haldane's model are well known and will not be repeated here.
%
Other variants of Haldane circuits can be constructed easily. For example, one can use capacitor 4-loops with 1-connection (nearest neighbor) and $\pm i$ (next nearest neighbor) connections. This will give rise to double Haldane model for $\Phi=\pi/4$. One can also add more wires to implement third nearest neighbor coupling $t_3$, which can give bands with Chern number $\pm 2$ \cite{sticlet2012geometrical}.

\subsection{Symmetries}
The 2D circuits B to E above share a common feature that the bands of interest are spin degenerate. More specifically, the Chern number for the spin up mode is the negative of the spin down mode. As a result, while the edge state of each spin is chiral, they propagate in opposite directions.
The key question then is how robust these edge states are, and what kind of perturbations can hybridize the spin up and down modes and gap out the edge spectrum.
The situation is thus similar to the quantum spin Hall effect. There the edge modes are protected by time reversal symmetry (applied to fermions),
scattering from nonmagnetic impurities cannot mix the spin up and down modes or open a gap.
For the circuits here, by design the $m$-connections do not couple different spin species and {each spin is separately conserved}.
More specifically, the phase factor $e^{im2\pi k/p}$ in Eq. \eqref{Lnk} is diagonal in the mode index $k$, and does not lead to spin mixing such as the $\sigma_{x,y}$ terms mentioned in section V.E.

To further clarify the symmetry behind the spin degeneracy, recall that the spin up and down modes correspond to a pair of degenerate modes,
$\varphi_k$ and $\varphi_{p-k}$ respectively, for instance, $k=1$ for the $p=3$ or $p=4$ capacitor loop. One can think of them as modes traveling
in the clockwise ($k$) or counterclockwise ($p-k$) direction along the loop. The eigen index $k$ follows from the $p$-fold rotational symmetry of the loop.
We can formally define an operator $\mathcal{P}$ that exchanges $k$ and $p-k$, i.e. reversing $k$ modulo $p$, and thus flips the spin. Now consider
a connection of type-$m$ between two loops. The spin up mode sees the connection as a phase shift $S_m=e^{im2\pi k/p}$, for the twist $j\rightarrow j+m$
is in the same screw direction of mode $k$. According to Eq. \eqref{Lnn+1},
the spin down mode sees a phase shift $S^*_m$, the complex conjugation of $S_m$, since the twist is in the opposite direction
of mode $p-k$. A loose optical analogy is the left and right circularly polarized light picking up different phases after a birefringent medium.
The circuit Lagrangian and equations of motion are invariant under $\mathcal{CP}$,
i.e. spin flip $\mathcal{P}$ followed by complex conjugation $\mathcal{C}$. The combined operation $\mathcal{CP}$ is a symmetry of
the circuit and commutes with $H$. It protects the edge state and plays the role of time reversal symmetry in quantum spin Hall effect.
Note however, the physical time reversal symmetry applies to any LC circuit, and by itself cannot explain the robustness of the edge states.
The related issues are discussed in Ref. \cite{albert2015topological}.

\section{Four dimensional topological circuits}

Finally, let us generalize the discussion to periodic lattices of arbitrary dimension $D$. Each node (e.g. a capacitor loop) is labelled
by a $D$-tuple, $\v{n}=(n_1,n_2,...,n_D)$, where the integer $n_i$ is the lattice coordinate along the $i$-th direction.
For example, on a 4D hypercube lattice,
%, also known as the tesseract, as part of the
%\begin{center}
% \includegraphics[width=0.4\textwidth]{fourcube.pdf}
%\end{center}
each node is denoted by four integers, $\v{n}=(n_x,n_y,n_z,n_w)$. Let $\hat{x},\hat{y},\hat{z},\hat{w}$ be
the four unit lattice vectors along the respective direction, e.g. $\hat{x}=(1,0,0,0)$. We stress that
the effective spatial dimension of a circuit network is determined by the connectivity among the nodes.
To realize a periodic lattice circuit, it is {\it not} necessary to literally lay out the circuit elements periodically in space. It only requires
the values of $L$ and $C$, and the wiring pattern, to repeat in $D$ independent directions.
There is no difficulty constructing a periodic 3D or 4D circuit on a breadboard.

The following example of 4D circuit illustrates these points. Start from a set of
identical square lattices (the green plaquettes in figure below) extending in the $x$ and $y$ direction.
Here each node is a capacitor 3-loop, and the solid line denotes the inductive $0$-connection.
Next connect these square lattices in the $z$ direction. The inductive
coupling depends on the $x$ coordinates: it is of type $m_x\equiv n_x$ modulo $3$. That is, three types of
connections, the $0$-connection (solid line), $1$-connection (dashed line), and $2$-connection (dotted line) alternate.
Finally, connect the square lattice in similar fashion in the $w$ direction with type $m_y\equiv n_y$ modulo $3$.
The result is the following circuit (the wires are shown only partially to avoid clutter).
\begin{center}
  \includegraphics[width=0.3\textwidth]{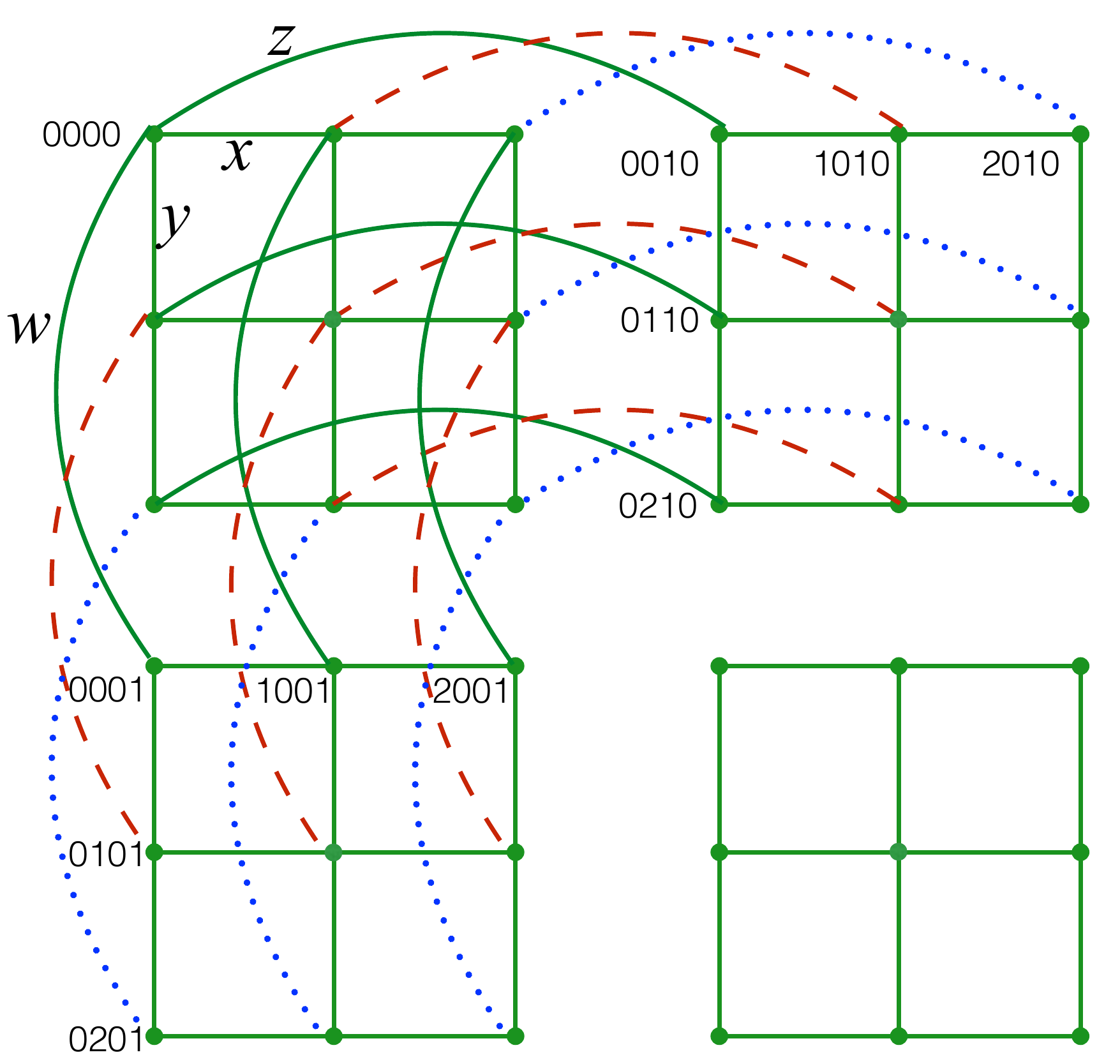}
\end{center}
Another perspective of the {same circuit} is as follows
\begin{center}
  \includegraphics[width=0.4\textwidth]{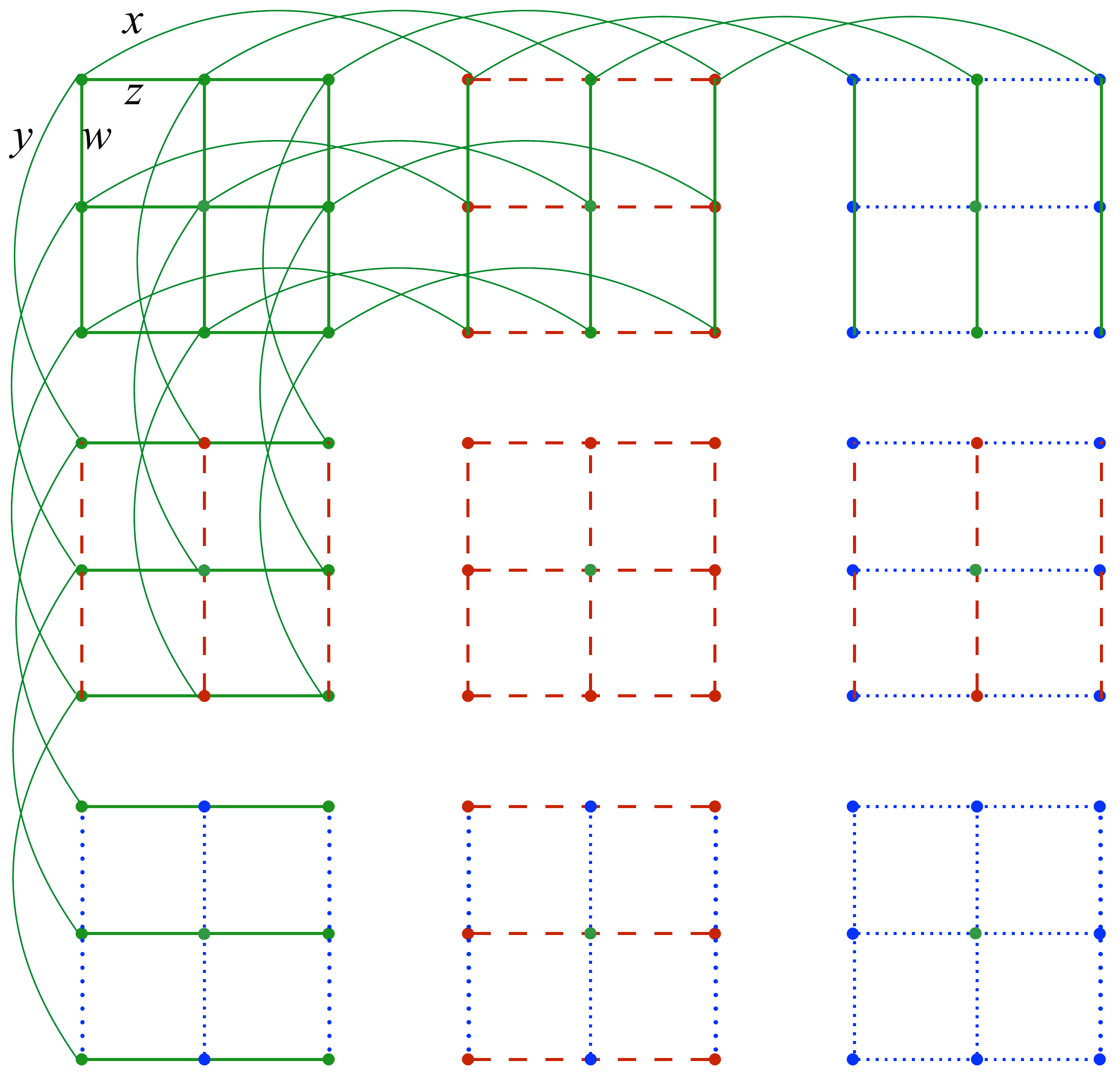}
\end{center}
High dimensional lattices like these are harder to imagine, but a moment's reflection will convince these are indeed
identical circuits. This 4D example also illustrates the advantage of having a simplified notation. Without it, the circuit
would appear to be a complete mess.

The circuit Lagrangian is actually easy to write down
\begin{flalign}
&\mathscr{L}=\frac{C}{2}\sum_{\v{n},j} [\dot{\phi}_{j}(\v{n})-\dot{\phi}_{j+1}(\v{n})]^2
 \\
&-\frac{1}{2L} \sum_{\v{n},j}[\phi_{j}(\v{n})- \phi_{j}(\v{n}+\hat{x})]^2+
[\phi_{j}(\v{n})- \phi_{j}(\v{n}+\hat{y})]^2  \nonumber \\
&+  [\phi_{j}(\v{n})- \phi_{j+ m_x}(\v{n}+\hat{z})]^2
+ [\phi_{j}(\v{n})- \phi_{j+ m_y }(\v{n}+\hat{w})]^2,\nonumber
\end{flalign}
where $j$ labels the flux variables inside each node, and $m_{x,y}$ are defined above as $n_x,n_y$ modulo 3.
After the Fourier transform Eq. \eqref{phi-to-varphi}, the equations of motion for mode $\ell=0,1,2$ decouple. For example,
the $\ell=1$ (spin up) mode is governed by equation
\begin{flalign}
(8-3\omega^2)\varphi_1(\v{n})&=\varphi_1(\v{n}+\hat{x})+\varphi_1(\v{n}-\hat{x}) \\
&+\varphi_1(\v{n}+\hat{y})+\varphi_1(\v{n}-\hat{y}) \nonumber \\
&+e^{i n_x\Phi }\varphi_1(\v{n}+\hat{z})+e^{-i n_x\Phi }\v\varphi_1(\v{n}-\hat{z}) \nonumber \\
&+e^{i n_y \Phi }\varphi_1(\v{n}+\hat{w})+e^{-i n_y\Phi }\v\varphi_1(\v{n}-\hat{w}), \nonumber
\end{flalign}
with $\Phi =2\pi/3$. The equation for the $\ell=2$ (spin down) mode is given by replacing $\Phi$ with $-\Phi$.
After Fourier transform $\v{n}\rightarrow (n_x,n_y,k_z,k_w)$, we find the eigenvalue problem factorizes,
i.e. $\varphi_1(n_x,n_y,k_z,k_w)=\varphi_{n_x}(k_z)\varphi_{n_y}(k_w)$, with $\varphi$ satisfying the Harper equation
\begin{align}
E_{xz} \varphi_{n_x}(k_z) &= e^{ik_x}\varphi_{n_x+1}(k_z) + e^{-ik_x}\varphi_{n_x-1}(k_z) \nonumber \\
& + 2\cos (n_x\Phi + k_z) \varphi_{n_x}(k_z), \\
E_{yw} \varphi_{n_y}(k_w) &= e^{ik_y}\varphi_{n_y+1}(k_w) + e^{-ik_y}\varphi_{n_y-1}(k_w) \nonumber \\
& + 2\cos (n_y\Phi + k_w) \varphi_{n_x}(k_w).
\end{align}
The eigen frequency is given by $E\equiv 8-3\omega^2=E_{xz}+E_{yw}$.
The solution to such Harper equation is well known from the study of the AH model \cite{satija2016butterfly}.
For flux $\Phi =2\pi/3$, it features three bands with first Chern number $c=1$, -2, and 1
respectively. From these, the bulk spectrum $\omega(k_x,k_y,k_z,k_w)$ can be obtained.
The figure below shows the density of states of the frequency spectrum.
\begin{center}
  \includegraphics[width=0.4\textwidth]{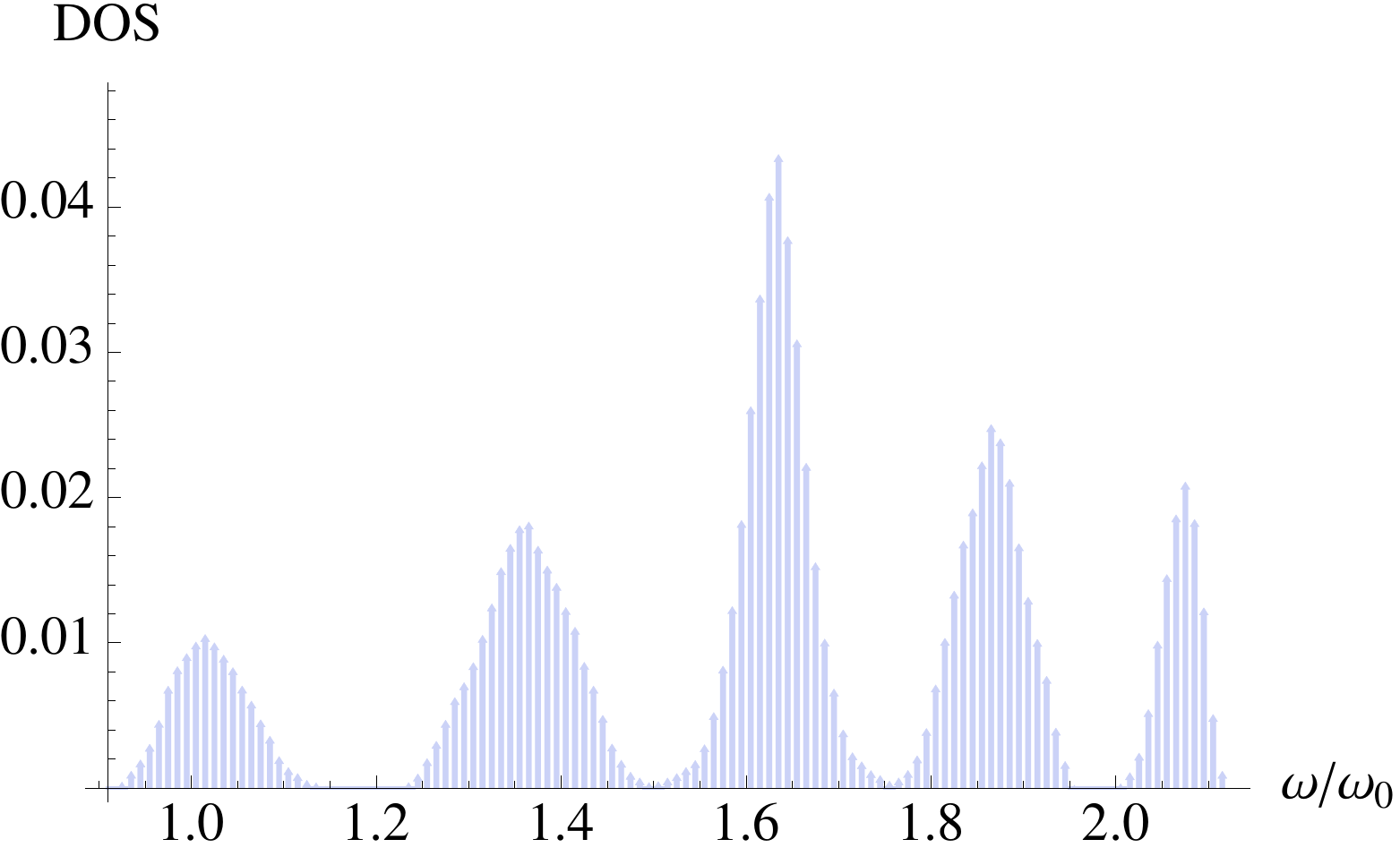}
\end{center}
The lowest and highest bands are separated from the rest by a well defined gap.

For the bottom band, define the second Chern number
\be
\mathscr{C}_2=\frac{1}{8\pi^2}\int_{BZ} F \wedge F
\ee
For our 4D circuit, it reduces to the product of the first Chern number
\be
\mathscr{C}_2=c_{xz} \times c_{yw}.
\ee
At flux 1/3, $c_{xz}= c_{yw}=1$, so we have $\mathscr{C}_2=1$. Note that 
similar 4D quantum Hall effect has been proposed \cite{price2015four} and realized in cold atoms \cite{lohse2018exploring}
and photonic waveguides \cite{zilberberg2018photonic}.

\section{Outlook}

In summary, we have described an assortment of topological LC circuits within a unified approach, building on the pioneering works of Refs. \cite{PhysRevX.5.021031,albert2015topological}. We have distilled the common design strategy behind these circuits (section \ref{blocks}) and shown how topological circuits in one, two, and four dimensions can be constructed based on this strategy. Along the way, we have introduced new building blocks (e.g. capacitor or inductor stars) and a simplified notation scheme. The notation is not just for convenience. It enforces modular design, i.e., constructing ever more complicated circuits using a few standard blocks and connections. 
We have also streamline the theoretical analysis by using the Lagrangians and $H$-matrix, $\mathscr{L}\rightarrow H(k) \rightarrow \omega(k)$. This general framework is applied repeatedly to treat all the circuit examples, and to clarify the connection between topological LC circuits and quantum lattice models as well as the role played by symmetries. This is facilitated by using a consistent language in terms of connection $A$, curvature $F$, pseudospins and synthetic gauge fields throughout. 

How useful are these topological LC circuits? First of all, they provide a convenient yet flexible platform to understand the topological properties of classical oscillations and waves.
Among the classical topological systems studied so far, LC circuits have the advantages of being easy to build and probe. More importantly, wire connection alone can offer versatile control over the coupling between the building blocks. For example, one can easily go beyond (effectively) three dimensions or nearest neighbor couplings. Any graph or network geometries such as tori, periodic or open boundaries, domains, defects can be realized. Some of these features would be much harder to achieve, for instance, in mechanical oscillators or photonic crystals. From an engineering point of view, a large class of electromagnetic metamaterials can be modeled by periodic circuits in the lumped-element description. An improved understanding of robust electromagnetic modes localized at domains, defects, or boundaries from the study of topological LC circuits may have implications for designing novel metamaterials, e.g., to fill the terahertz gap.

The work presented here points to a few possible directions for future work on topological circuits. One can imagine all kinds of ``crystals" (circuits) made from ``molecules" (building blocks) of the ``atoms" (L and C). So far, we have only considered simple unit cells, lattice geometries, and wiring patterns. Special emphasis is placed on exploiting the discrete translation and permutation symmetry of the loops and stars by braiding ($m$-shift in particular). More interesting band structures may result from nontrivial point group or nonsymmorphic space group symmetries.  We have not included transformers, gyrators or other circuits elements in our design. The couplings between the pseudospin degree of freedom or the analogs of spin-orbit coupling have not been included either. Finally, classical linear LC circuits are  the first step toward understanding topological quantum circuits involving Josephson junctions where the flux variable $\phi$ is quantized and conjugate to the number operator, giving rise to the possibility of bosonic symmetry protected topological phases.

\begin{acknowledgments}
This research is supported by NSF PHYS-1707484 and AFOSR Grant No. FA9550-16-1-0006. 
\end{acknowledgments}
%{\bf Acknowledgments}. This research is supported by NSF PHYS-1707484 and AFOSR Grant No. FA9550-16-1-0006.

%\bibliographystyle{elsarticle-num} 
\bibliography{refs}
\end{document}